\documentclass{amsart}[12pt]
\usepackage{amsmath}
\usepackage{amstext}
\usepackage{amssymb}
\usepackage{latexsym}
\numberwithin{equation}{section}

\advance\textheight3truecm
\advance\hoffset-2truecm
\advance\voffset-1.7truecm
\newtheorem{theorem}{Theorem}[section]
\newtheorem{proposition}[theorem]{Proposition}
\newtheorem{lemma}[theorem]{Lemma}
\theoremstyle{definition}
\newtheorem{definition}[theorem]{Definition}

\newtheorem{corollary}{Corollary}[section]
\theoremstyle{remark}
\newtheorem{remark}[theorem]{Remark}
\numberwithin{equation}{section}

\begin{document}
\title{Gabor-type Frames from Generalized
        Weyl-Heisenberg Groups}

\author{G. Honnouvo}
\email{g$\_$honnouvo@yahoo.fr}
\author{S. Twareque Ali}
\email{stali@mathstat.concordia.ca}
\address{Department of Mathematics and Statistics, Concordia
University,
   7141 Sherbrooke Street West, Montreal, Qu\'ebec, CANADA H4B 1R6}

\date{\today}

\begin{abstract}We present in this paper a construction for Gabor-type
frames
built out of generalized Weyl-Heisenberg groups. These latter are
obtained
via central extensions of groups which are direct products of  locally
compact abelian groups and their duals. Our results generalize many of
the results,
appearing in the literature, on frames built out of the Schr\"odinger
representation
of the standard Weyl-Heisenberg group. In particular, we obtain a
generalization of the result in  \cite{PO}, in which the product
$ab$ determines whether it is possible for the Gabor system
$\{E_{mb}T_{na}g  \}_{m,n\in \mathbb Z}$ to be a frame for $L^2(\mathbb
R)$.
As a particular example of the theory, we study in some detail the case
of
the generalized Weyl-Heisenberg group built out of the $d$-dimensional
torus.
In the same spirit we also construct generalized shift-invariant
systems.
\end{abstract}

\maketitle
\Large
\section{Introduction}
Recently, a generalization of the Weyl-Heisenberg group has been
presented in \cite{K1,K}. Such a generalized Weyl-Heisenberg group
is the central extension of the direct product of a locally
compact abelian group $G$ with its dual group $\hat{G}$. By
analogy with the standard Weyl-Heisenberg group, it is then
possible to construct Schr\"odinger-type representations in these
general situations, which are again continuous, unitary and
irreducible. Since a Gabor system can be considered as the orbit
of a discrete subset of the  Weyl-Heisenberg group, under the
Schr\"odinger action, this paper presents a generalization of such
a system using a discrete subset of the generalized
Weyl-Heisenberg group. There exists an abundance of frame-related
results, in the literature, for Gabor frames. A good sampling of
these may be found, for example, in \cite{PO,DG,FS,GR,HW,RS}. This
paper presents extensions of several of these results, in
particular those dealing with the boundedness and invertibility of
the frame operator, to generalized Gabor systems. Specifically, we refer to
Theorems \ref{Th4}, \ref{Th5}, \ref{Th6} and \ref{Th11} below.
We ought to also mention that generalized Weyl-Heisenberg groups have
also been looked at for other and related studies. A good reference
is the work edited by
G. Feichtinger and Werner Kozek \cite{FS} and in particular,  Chapter 7
of that book.

The rest of this paper is organized as follows:
Section 2 sets out  some definitions and a few known results related
to
Gabor frames. Section 3 presents the definition of the generalized
Weyl-Heisenberg group and generalizations, in this setting,  of some
results mentioned in
the previous section. Section 4 is devoted to the study of  a special
case:
the generalized Weyl-Heisenberg group associated to the torus $\mathbb
T^d$. In
Theorem \ref{Th5} we work out the analogue on
$L^2(\mathbb T^d)$,
for this group, of the (by now) standard result that the product $ab$
puts a condition on the system $\{E_{mb}T_{na}g  \}_{m,n\in \mathbb Z}$
to be a frame
for $L^2(\mathbb R)$. This result is further generalized, to any LCA group,
in Theorem \ref{Th6}. Finally, we define a generalized shift-invariant
system on $L^2(G)$
and present some results associated to it.

\section{Preliminaries}\label{sec:prelim}
We begin by giving the central definitions and some necessary and
sufficient
conditions for a standard  Gabor system to be a frame.

\subsection{Some definitions}

A sequence $\{ f_k \}_{k=1}^\infty$ of elements in a Hilbert space
$\mathcal H$ is
called a Bessel sequence if there exists a constant $B>0$ such that
$$\sum_{k=1}^\infty\mid\langle f,f_k\rangle\mid^2\leq B\parallel
f\parallel^2,\quad
\forall f\in\mathcal H.$$

A Riesz basis for $\mathcal H$ is a family of the form $\{ Ue_k
\}_{k=1}^\infty$,
where  $\{ e_k \}_{k=1}^\infty$ is an orthonormal basis for $\mathcal
H$ and
$U:\mathcal H\longrightarrow \mathcal H$ is a bounded bijective
operator.

\begin{enumerate}
\item[(i)]

A sequence $\{ f_k \}_{k=1}^\infty$ of elements in a Hilbert space
$\mathcal H$
is  a frame for $\mathcal H$ if there exist constants $A,B>0$ such
that
\begin{equation*}
A\parallel f\parallel^2\leq\sum_{k=1}^\infty\mid\langle
f,f_k\rangle\mid^2\leq
B\parallel f\parallel^2,\quad \forall f\in \mathcal H.
\end{equation*}
The numbers $A$ and $B$ are called frame bounds.
\item[(ii)]
A frame is tight if we can choose $A=B$ as frame bounds.
\end{enumerate}

The following Lemma will be useful in the sequel.
\begin{lemma}\label{RB}
Let $\{ f_k\}_{k=1}^\infty,$ be a frame. Then the following are
equivalent:
\begin{enumerate}
\item[(i)]
$\{ f_k\}_{k=1}^\infty$ is tight.
\item[(ii)]
$\{ f_k\}_{k=1}^\infty$ has a dual of the form $g_k=Cf_k,$ for some
constant $C>0.$
\end{enumerate}
\end{lemma}

\subsection{Weyl-Heisenberg frame}

Let $x,w$ be real numbers. The unitary operators defined on
$L^2(\mathbb R)$ by
$T_xf(y)= f(y-x)$, and $E_wf(y)=e^{2\pi w.y}f(y),$ are called
translation and
modulation operators, respectively.

A Weyl-Heisenberg frame, or synonymously, a Gabor frame is a frame for
$L^2(\mathbb R)$ of the form $\{E_{mb}T_{na}g \}_{m,n\in
\mathbb Z},$ where $a,b>0$ and $g\in L^2(\mathbb R)$ is a fixed
function.
Explicitly,
 $$E_{mb}T_{na}g(x)=e^{2\pi mbx}g(x-na).$$
The function $g$ is called the window function or the frame generator.
For an exhaustive list of papers dealing with such frames we refer to
the
monograph \cite{FS}.

  Our main results in this paper will consist of generalizations of
the following four theorems for standard Gabor or Weyl-Heisenberg
frames.

\begin{theorem}\label{Th0}
Let $g\in L^2(\mathbb R)$ and $a,b>0$ be given. Then the following
holds:
\begin{eqnarray}
&&(i)\:\mbox{If}\quad ab>1,\: then \:\{E_{mb}T_{na}g \}_{m,n\in \mathbb
Z}\:\:
\mbox{is not a frame for}\: L^2(\mathbb R).\nonumber\\
&&(ii)\:\mbox{If}\quad\{E_{mb}T_{na}g \}_{m,n\in \mathbb Z}\:\:
\mbox{is a frame},\:
then\nonumber\\
&&ab=1 \Leftrightarrow  \{E_{mb}T_{na}g \}_{m,n\in \mathbb Z}\:\:
\mbox{is a Riesz basis.}\nonumber
\end{eqnarray}
\end{theorem}

In \cite{Ri}, there is the following stronger result than $(i):$ when
$ab>1,$ the family
$ \{E_{mb}T_{na}g \}_{m,n\in \mathbb Z}$ can not even be complete in
$L^2(\mathbb R).$
The assumption $ab\leq 1$ is not enough for  $\{E_{mb}T_{na}g
\}_{m,n\in \mathbb Z}$
to be a frame, even if $g\neq 0.$ For example, if $a\in
[\frac{1}{2},1[,$ the set of
functions $ \{E_{m}T_{na}\chi_{[0,\frac{1}{2}]} \}_{m,n\in \mathbb Z}$
is not
complete in $L^2(\mathbb R)$ and cannot form a frame.

 Proposition 8.3.2 in \cite{O} gives a necessary condition for a
Gabor
system to be a frame.
Sufficient conditions for $\{E_{mb}T_{na}g \}_{m,n\in \mathbb Z}$ to be
a
frame for $L^2(\mathbb R)$ have been known since 1988, the basic
insight being
provided by Daubechies \cite{DI}.
 A slight improvement was proved in \cite{HW}.
Later, Ron and Shen \cite{RS} were able to give a complete
characterization of
Gabor frames, spelled out in the next theorem. Given $g\in L^2(\mathbb
R),$
consider the matrix-valued function
\begin{equation}\label{ST}
M(x)=\left\{ g\left(x-na-m/b\right)  \right\}_{m,n\in \mathbb Z}.
\end{equation}
The matix $M(x)M^*(x)$ is positive.

\begin{theorem}\label{ST1}
$\{E_{mb}T_{na}g \}_{m,n\in \mathbb Z}$ is a frame for $L^2(\mathbb R)$
with
bounds $A,B$ if and only if
\begin{equation}
bA\mathbb I\leq M(x)M^*(x)  \leq bB\mathbb I,\: a.e.,
\end{equation}
where $\mathbb I$ is the identity operator on  $\ell^2(\mathbb Z).$
\end{theorem}
This theorem is a special case of the following characterization
\cite{O}
of a shift-invariant system to be a frame.

  Recall that if  $\{ g_m \}_{m\in I}$ is a collection of functions in
$L^2 (\mathbb R)$,  the
shift-invariant system generated by $\{ g_m \}_{m\in I}$  and some
$a \in \mathbb R$ is the collection
of functions $\{ g_m (.-na)\}_{m\in In\in \mathbb Z}.$   Usually we
will set
$I=\mathbb Z.$

Given a shift-invariant system $\{g_m (.-na)\}_{n,m\in \mathbb Z}$ for
$L^2(\mathbb R)$,
define the matrix-valued function $H(\nu), \; \nu \in \mathbb R$, by
\begin{eqnarray}
H(\nu)&=&\left(\hat{g}_m(\nu-k/a)  \right)_{k,m\in\mathbb Z},\quad
a.e.\; ,
\end{eqnarray}
$\hat{g}$ denoting the Fourier transform of $g$.
The following theorem then contains a generalization of Theorem
\ref{ST1} to any
shift-invariant system $\{g_m (.-na)\}_{n,m\in \mathbb Z}$ for
$L^2(\mathbb R)$:

\begin{theorem}\label{SIS} With the above setting, the following hold:
\begin{enumerate}
\item[(i)]
$\{ g_{n,m}\}$ is a Bessel sequence with upper bound $B$ if and only
if
$H(\nu)$, for almost all  $\nu,$ defines a bounded operator on
$l^2(\mathbb Z)$
of norm at most $\sqrt{a B}.$
\item[(ii)]
$\{ g_{n,m}\}$ is a frame for $L^2(\mathbb R)$ with frame bounds
$A,\:B$ if and only if
\begin{equation}
aA\mathbb I\leq H(\nu)H^*(\nu)\leq a B\mathbb I,\quad a.e.\:\nu.
\end{equation}
\item[(iii)]
$\{ g_{n,m} \}$ is a tight frame for $L^2(\mathbb R)$ if and only if
there is a
constant $c>0$ such that
\begin{equation}\label{si}
\sum_{m\in \mathbb Z}\overline{\hat g_m(\nu)}\hat
g_m(\nu+k/a)=c\delta_{k,0},
\quad \:k\in\mathbb Z,\:a.e.\:\nu.
\end{equation}
In case $(\ref{si})$  is satisfied, the frame bound is $A=c/a.$
\item[(iv)]
Two shift-invariant systems $\{ g_{n,m} \}$ and $\{h_{n,m} \}$, which
form
Bessel sequences, are dual frames if and only if
\begin{equation}\label{f5}
\sum_{m\in \mathbb Z}\overline{\hat g_m(\nu)}\hat
h_m(\nu+k/a)=a\delta_{k,0},
\quad  k\in\mathbb Z,\:a.e.\:\nu.
\end{equation}
\end{enumerate}
\end{theorem}

Theorem \ref{ST1} is difficult to apply. However, the condition on the
matrix
$M(x)M^*(x)$ is in particular satisfied if it is diagonal dominant.
This leads to
a sufficient condition \cite{PO} for $\{E_{mb}T_{na}g \}_{m,n\in
\mathbb Z}$ to be a frame
for $L^2(\mathbb R)$.

\begin{theorem}\label{Th3}
Let $g\in L^2(\mathbb R)$ and $a,b>0$ and suppose that
\begin{equation}
 B:=\frac{1}{b}\sup_{x\in [0,a]}\sum_{k\in \mathbb Z}\mid \sum_{n\in
 \mathbb Z}g(x-na)\overline {g(x-na-k/b)}\mid< \infty.
\end{equation}
Then $\{E_{mb}T_{na}g \}_{m,n\in \mathbb Z}$ is a Bessel sequence with
upper
frame bound $B$. If also
\begin{equation}
A:=\frac{1}{b}\inf_{x\in [0,a]}\left[\sum_{n\in \mathbb Z}\mid g(x-na)
\mid^2-\sum_{0\neq k\in \mathbb Z}| \sum_{n\in \mathbb Z} g(x-na)
\overline {g(x-na-k/b)}\mid\ \right]>0.
\end{equation}
Then $\{E_{mb}T_{na}g \}_{m,n\in \mathbb Z}$ is a frame for
$L^2(\mathbb R)$
with bounds $A,B.$
\end{theorem}

\section{Generalized Weyl-Heisenberg Group}

\begin{definition}
Let $ G$ be a locally compact abelian (LCA) group, $\hat G$ its dual
group,
$\mu$ and $\nu$ their Haar
measures, respectively. Let $\mathbb T$ be the unit circle and  put
$H_{G}=
G\times\hat{ G }\times \mathbb T.$  For $(g_1,w_1,z_1)$ and
$(g_2,w_2,z_2)$ in
$H_{ G}$, define the following composition:
\begin{eqnarray}
(g_1,w_1,z_1).(g_2,w_2,z_2)=\big(g_1g_2,w_1w_2,z_1z_2w_2(g_1)\big).
\end{eqnarray}
$H_{ G}$ is closed under this action, which is also  associative.
Equipped with
this product,  $H_{ G}$ is a group, called the
generalized Weyl-Heisenberg group, associated
with ${G}.$ This group is nonabelian, locally compact, and unimodular
\cite{SK}, with
invariant measure $d\mu d\nu d\theta$ (where $z = e^{i\theta}$).
\end{definition}
A uniform lattice in  $ G$ is a discrete subgroup $ K$ of $ G$ such
that
$ G/{K}$ is compact.
For a uniform lattice $ K$ in $ G,$  $Ann( K),$ denotes the annihilator
of
$ K$, {\em i.e.\/,}  $Ann( K):=\{\gamma\in \hat{ G}\; :\; \gamma(k)=1,
\forall \:k\in  K  \}.$

By Lemma 24.5 of \cite{ROS}, we know that $Ann( K)=\widehat{ G/K}$, so that
$Ann( K)$ is a
discrete subgroup of $\hat G.$

Let $\pi:H_{ G}\longrightarrow U\Big(L^2({ G}) \Big)$ be the
Schr\"odinger
representation of $H_{ G}$, which is a unitary, irreducible
representation, given
explicitly by
\begin{equation}
\Big(\pi(x,\gamma,z)g  \Big)(t)=z\gamma(t)g(tx^{-1}),
\end{equation}
for all $(x,\gamma,z)\in H_{ G}$ and almost all $g\in L^2({ G})$.
In \cite{K1}, frames of $L^2(G)$ of the type
\begin{eqnarray}\label{r1}
\left\{\Theta^g_{(k,\gamma)}=\Big(\pi(k,\gamma,1)g
\Big):\:(k,\gamma)\in K\times
Ann( K) \right\},
\end{eqnarray}
where $K$ is a  uniform lattice in $G$, have been studied.
In this case, taking $G=\mathbb R$ and $K= a\mathbb Z$ and  defining
the
dual pairing in the usual way:
\begin{eqnarray}
\xi(x)=e^{2\pi ix\xi},
\end{eqnarray}
we obtain $Ann(K)= \displaystyle\frac{1}{a}\mathbb Z.$ Thus, the Gabor
system defined by
(\ref{r1}) is
\begin{eqnarray}
\left\{ e^{\frac{2\pi imx}{a}}g(x-na)\right\}_{m,n\in \mathbb Z},
\end{eqnarray}
which is a particular case ($ab=1$) of the standard Gabor system:
\begin{eqnarray}\label{r2}
\left\{ e^{2\pi imbx}g(x-na)\right\}_{m,n\in \mathbb Z}.
\end{eqnarray}

    In this paper, we study  frames for $L^2(G)$ of the form
\begin{eqnarray}
\left\{\Theta^g_{(k_1,\gamma_2)}=
\Big(\pi(k_1,\gamma_2,1)g\Big):\:(k_1,\gamma_2)
\in K_1\times Ann( K_2) \right\},
\end{eqnarray}
where $K_1$ and $K_2$ are two lattices in $G$.
Such a  frame is clearly a generalization of a Gabor frame,  because if
we
take $K_1= a\mathbb Z$
and $K_2= \displaystyle\frac{1}{b}\mathbb Z$ and use the same dual
pairing as above,
we get exactly the standard Gabor system (\ref{r2}).

\begin{definition} The set defined by
\begin{equation}
\left\{\Theta^g_{(k_1,\gamma_2)}:\:(k_1,\gamma_2)\in K_1\times Ann(
K_2) \right\}
\end{equation}
will be called the generalized Gabor system for $L^2(G)$ associated to
the uniform
lattices $K_1$ and  $K_2$, and the window function $g.$
\end{definition}

The following Lemmata are essential for this work:

\begin{lemma}\label{Lem1} Let $f,g\in L^2(G).$ and $ K_1$ and  $ K_2$
be two
uniform lattices in $G.$
Then, for any $k_2\in  K_2,$ the series
\begin{equation}
\sum_{k_1\in K_1}f(xk_1^{-1})\overline{g(xk_1^{-1}k_2^{-1})}
\end{equation}
converges absolutely for almost all $x\in G$ and the function
\begin{equation}
x\longmapsto \sum_{k_1\in K_1}\mid
f(xk_1^{-1})\overline{g(xk_1^{-1}k_2^{-1})}
\mid \; \in\;  L^1(G/{ K_1}),
\end{equation}
\end{lemma}

\begin{proof}
Since $f\in L^2(G)$ and $T_{k_2}g\in L^2(G)$, then $f.T_{k_2}g\in
L^1(G).$ But,
\begin{equation}
\int_{G/{K_1}}\sum_{k_1\in K_1}\mid
f(xk_1^{-1})\overline{g(xk_1^{-1}k_2^{-1})}
\mid dx=\int_{G}\mid f(x)\overline{g(xk_2^{-1})}\mid dx <\infty,
\end{equation}
by H\"older's inequality. This implies that
$\sum_{k_1\in K_1}\mid f(xk_1^{-1})\overline{g(xk_1^{-1}k_2^{-1})}\mid
<\infty\:, a.e.$
\end{proof}

\begin{lemma}\label{Lem2} Let $f,g\in L^2(G)$ and  $ K_1$, $K_2$ be two
uniform
lattices in $G.$
For a fixed $k_1\in  K_1,$ consider the function $F_{k_1}\in L^1(G/{
K_2}),$ defined by
\begin{eqnarray}
F_{k_1}(x)=\sum_{k_2\in
K_2}f(xk_2^{-1})\overline{g(xk_1^{-1}k_2^{-1})}.
\end{eqnarray}
Then, for any $(k_1,\gamma_2)\in K_1\times Ann({ K_2}),$ we have
\begin{eqnarray}
\langle f|\Theta^g_{(k_1,\gamma_2)}\rangle =\int_{G/{K_2}}
\overline{\gamma_2(x)}F_{k_1}(x)dx.
\end{eqnarray}
\end{lemma}

\begin{proof}
\begin{eqnarray}
\langle f|\Theta^g_{(k_1,\gamma_2)}\rangle &=&\int_G
\overline{\gamma_2(x)}f(x)
\overline{g(xk_1^{-1})}dx\nonumber\\
&=&\int_{G/{ K_2}}\sum_{k_2\in  K_2}
\overline{\gamma_2(xk_2^{-1})}f(xk_2^{-1})
\overline{g(xk_1^{-1}k_2^{-1})}dx\nonumber\\
&=&\int_{G/{K_2}}\overline{\gamma_2(x)}\sum_{k_2\in  K_2}f(xk_2^{-1})
\overline{g(xk_1^{-1}k_2^{-1})}dx\nonumber\\
&=&\int_{G/{ K_2}}\overline{\gamma_2(x)}F_{k_1}(x)dx.
\end{eqnarray}
\end{proof}

  Let us now present the generalization of the WH-frame identity
(see \cite{O}, Lemma 8.4.3) for an arbitrary LCA group $G$. Our generalization
appears in Lemma \ref{Lem3} below. Consider the function $H_{1_G}$
defined by
\begin{equation}\label{f15}
   H_{1_G}(x)=\sum_{k_1\in  K_1}\mid g(xk_1^{-1})\mid^2\; .
\end{equation}
\begin{lemma}\label{Lem3}
Let $f,g\in L^2(G)$ and  $K_1, K_2$ be two uniform lattices in $G.$
Suppose
that $f$ is a bounded measurable function with compact support and that
the
function $H_{1_G}$ defined by (\ref{f15}) is bounded. Then
\begin{eqnarray}
&&\sum_{k_1\in K_1} \sum_{\gamma_2\in Ann( K_2)} \mid\langle
f|\Theta^g_{(k_1,\gamma_2)}\rangle\mid^2\\
&=&\mid G/{ K_2}\mid\int_G \mid f(x)\mid^2\sum_{k_1\in  K_1}\mid
g(xk_1^{-1})
\mid^2dx\nonumber\\
&+&\mid G/{ K_2}\mid\sum_{1_G\neq k_2\in
K_2}\int_G\overline{f(x)}f(xk_2^{-1})
\sum_{k_1\in  K_1}
g(xk_1^{-1})\overline{g(xk_1^{-1}k_2^{-1})}dx\nonumber\; ,
\end{eqnarray}
where $\mid G/{ K_2}\mid$ denotes the measure of $G/{ K_2}$.
\end{lemma}
\begin{proof}
From Theorem 4.26 in \cite{Fol}, we conclude that the set of functions
$\{ \mid G/{ K_2}\mid^{-\frac{1}{2}}\gamma_2 \}_{\gamma_2\in Ann(
K_2)},$ is
an orthonormal basis for $L^2(G/{ K_2}).$ \mbox{By Parseval's theorem,
we have}
\begin{eqnarray}
\sum_{\gamma_2\in Ann( K_2)}\mid \int_{G/{ K_2}}\overline{\gamma_2(x)}
F_{k_1}(x)dx\mid^2= \mid G/{ K_2}\mid\int_{G/{ K_2}}\mid F_{k_1}(x)
\mid^2 dx.
\end{eqnarray}
which implies that
\begin{eqnarray}
&&\sum_{k_1\in  K_1} \sum_{\gamma_2\in Ann( K_2)} \mid\langle
f|\Theta^g_{(k_1,\gamma_2)}\rangle\mid^2\\
&=&\sum_{k_1\in  K_1} \sum_{\gamma_2\in Ann(K_2)}\mid \int_{G/{K_2}}
\overline{\gamma_2(x)}F_{k_1}(x)dx\mid^2\nonumber\\
&=&\mid G/{ K_2}\mid\sum_{k_1\in K_1}\int_{G/{ K_2}}\mid
F_{k_1}(x)\mid^2 dx
\nonumber\\
&=&\mid G/{ K_2}\mid\sum_{k_1\in  K_1}\int_{G/{
K_2}}{F_{k_1}(x)}\sum_{k_2\in  K_2}
\overline{{f(xk_2^{-1})}} {g(xk_2^{-1}k_1^{-1})}dx\nonumber\\
&=&\mid G/{ K_2}\mid\sum_{k_1\in K_1}\int_{G}{F_{k_1}(x)}
\overline{{f(x)}}{g(xk_1^{-1})}dx\nonumber\\
&=&\mid G/{ K_2}\mid\sum_{k_1\in
K_1}\int_G\overline{{f(x)}}{g(xk_1^{-1})}
\sum_{k_2\in  K_2}{f(xk_2^{-1})}\overline{{g(xk_2^{-1}k_1^{-1})}}dx
\nonumber\\
&=&\mid G/{ K_2}\mid\int_G\mid f(x)\mid^2\sum_{k_1\in K_1} \mid
g(xk_1^{-1})
\mid^2dx\nonumber\\
&+&\mid G/{ K_2}\mid\sum_{1_G\neq k_2\in
K_2}\int_G\overline{f(x)}{f(xk_2^{-1})}
\sum_{k_1\in
K_1}{g(xk_1^{-1})}\overline{{g(xk_1^{-1}k_2^{-1})}}dx.\nonumber
\end{eqnarray}
\end{proof}

 The following is a generalization of Theorem \ref{Th3} to any LCA
group $G.$

\begin{theorem}\label{Th4} Let $K_1$ and $ K_2$ be two uniform lattices
of the
LCA group $G$. Let $g\in L^2(G)$ such that:
\begin{equation}
 B:=\mid G/{ K_2}\mid \sup_{x\in G/{K_1}}\sum_{k_2\in  K_2}\mid
\sum_{k_1\in
 K_1}g(xk_1^{-1})\overline {g(xk_1^{-1}k_2^{-1})}\mid < \infty.
\end{equation}
Then $\{\Theta^g_{k_1,\gamma_2 } \}_{(k_1,\gamma_2)\in K_1\times Ann(
K_2)}$ is
a Bessel sequence with upper frame bound $B$. If also
\begin{equation}
A:=\mid G/{ K_2}\mid \inf_{x\in G/{ K_1}}\left[\sum_{k_1\in  K_1}\mid
g(xk_1^{-1})\mid^2-\sum_{1_G\neq k_2\in  K_2}| \sum_{k_1\in  K_1}
g(xk_1^{-1})\overline {g(xk_1^{-1}k_2^{-1})}\mid\ \right]>0\; ,
\end{equation}
then $\{\Theta^g_{k_1,\gamma_2 } \}_{(k_1,\gamma_2)\in K_1\times Ann(
K_2)}$
is a frame for $L^2(G)$ with bounds $A,B.$
\end{theorem}

\begin{proof}

For $k_2\in K_2,$ fixed, define the function $H_{k_2}$ by

$H_{k_2}(x)=\sum_{k_1\in  K_1} T_{k_1}g(x)\overline{T_{k_1k_2}g(x)}.$
We have:
\begin{eqnarray}
\sum_{1_G\neq k_2\in  K_2} \mid T_{k_2^{-1}}H_{k_2}(x) \mid &=&
\sum_{1_G\neq k_2\in  K_2} \mid T_{k_2^{-1}}\sum_{k_1\in K_1}
T_{k_1}g(x)
\overline{T_{k_1k_2}g(x)} \mid\nonumber\\
&=&\sum_{1_G\neq k_2\in K_2} \mid \sum_{k_1\in  K_1}
T_{k_1k_2^{-1}}g(x)
\overline{T_{k_1}g(x)} \mid.
\end{eqnarray}
Replacing $k_2$ by $k_2^{-1},$ we have

\begin{eqnarray}
\sum_{1_G\neq k_2\in K_2} \mid T_{k_2^{-1}}H_{k_2}(x) \mid &=&
\sum_{1_G\neq k_2\in  K_2} \mid T_{k_2^{-1}}\sum_{k_1\in  K_1}
T_{k_1}g(x)
\overline{T_{k_1k_2}g(x)} \mid\nonumber\\
&=&\sum_{1_G\neq k_2\in K_2} \mid \sum_{k_1\in K_1}
T_{k_1k_2^{-1}}g(x)
\overline{T_{k_1}g(x)} \mid.\nonumber\\
&=&\sum_{1_G\neq k_2\in K_2} \mid \sum_{k_1\in K_1} T_{k_1}g(x)
\overline{T_{k_1k_2}g(x)} \mid.\nonumber\\
&=&\sum_{1_G\neq k_2\in K_2} \mid H_{k_2}(x) \mid.
\end{eqnarray}
Thus,
\begin{eqnarray}
&&\mid \sum_{1_G\neq k_2\in K_2} \int_G\overline{f(x)}f(xk_2^{-1})
\sum_{k_1\in K_1} g(xk_1^{-1})\overline{g(xk_1^{-1}k_2^{-1})}dx \mid
\\
&=&\mid \sum_{1_G\neq k_2\in K_2}
\int_G\overline{f(x)}T_{k_2}f(x)H_{k_2}(x) dx
\mid,\nonumber\\
&\leq&\sum_{1_G\neq k_2\in K_2} \int_G \mid f(x)\mid . \mid
T_{k_2}f(x)\mid .
\mid H_{k_2}(x)\mid dx\nonumber\\
&& \mbox{(using the Cauchy-Schwarz inquality twice: with
respect}\nonumber\\
&& \mbox{to the integral and  the sum)}\nonumber\\
&\leq& \int_G \mid f(x)\mid^2 . \sum_{1_G\neq k_2\in K_2} \mid
H_{k_2}(x)
\mid dx.\nonumber
\end{eqnarray}

By Lemma \ref{Lem3}, we have
\begin{eqnarray}
&&\sum_{k_1\in K_1} \sum_{\gamma_2\in Ann(K_2)}\mid \langle
f|\Theta^g_{(k_1,\gamma_2)}\rangle\mid^2\nonumber\\
&\leq&\mid G/{ K_2}\mid\int_G\left( \mid f(x)\mid^2\bigg[H_{1_G}(x) +
\sum_{1_G\neq k_2\in  K_2} \mid \sum_{k_1\in K_1} T_{k_1}g(x)
\overline{T_{k_1k_2}g(x)} \mid\bigg]\right)dx\nonumber\\
&=&\mid G/{ K_2}\mid\int_G dx\mid f(x)\mid^2\sum_{ k_2\in  K_2}
\mid \sum_{k_1\in  K_1} T_{k_1}g(x)\overline{T_{k_1k_2}g(x)}
\mid\nonumber\\
&\leq&B\mid\mid f\mid\mid^2.\nonumber
\end{eqnarray}
Also, we have
\begin{eqnarray}
&&\sum_{k_1\in K_1} \sum_{\gamma_2\in Ann(K_2)}\mid
\langle f|\Theta^g_{(k_1,\gamma_2)}\rangle\mid^2\nonumber\\
&\geq&\mid G/{ K_2}\mid\int_G\left( \mid f(x)\mid^2\bigg[H_{1_G}(x)-
\sum_{1_G\neq k_2\in \mathcal K_2}  \mid \sum_{k_1\in  K_1}
T_{k_1}g(x)
\overline{T_{k_1k_2}g(x)} \mid\bigg]\right)dx\nonumber\\
&\geq&A\parallel f\parallel^2.\nonumber
\end{eqnarray}
Since the frame conditions hold for all $f$ in a dense subspace of
$L^2(G)$,
it is true for any element of $L^2(G)$.
\end{proof}

\begin{remark}
The above result is more general than, and is in fact an extension to
other classes of groups,
of the results in \cite{PO,HW}. By taking $G=\mathbb R$, $ K_1=
a\mathbb Z,$
and $ K_2=\displaystyle\frac{1}{b}\mathbb Z,$ we recover Theorem
\ref{Th3}.
\end{remark}

\section{Frames on the torus $\mathbb T^d$}
Let $G=\mathbb T^d$ be the torus in $d$ dimensions.
Let $N_i,M_i\in \mathbb N^*, \; i=1,2, ...,d$, be $2d$ positive integers.
For simplicity, we adopt the following notation in this section:

Let $\underline{\mathfrak n}=(n_1,...,n_d),\; \underline{\mathfrak
N}=(N_1,...,N_d)$
and $\underline{\mathfrak M}=(M_1,...,M_d)$.
Set $\left(\underline{\mathfrak n},\underline{\mathfrak N}
\right)=\left(\frac{n_1}{N_1},\frac{n_2}{N_2},...,\frac{n_d}{N_d}\right)$
and $\left(\underline{\mathfrak m},\underline{\mathfrak M}  \right)=
\left(\frac{m_1}{M_1},\frac{m_2}{M_2},...,\frac{m_d}{M_d}\right)$
and consider the following two uniform lattices in $\mathbb T^d$:

\begin{eqnarray}
&&\mathcal K_1^{\underline{\mathfrak N}}=\left
\{\left(\underline{\mathfrak n},
\underline{\mathfrak N}
\right):n_i=0,1,...,N_i-1,\:\mbox{for}\:\:i=1,...,d  \right \}\\
&\mbox{\rm and} &\nonumber\\
&&\mathcal K_2^{\underline{\mathfrak M}}=\left
\{\left(\underline{\mathfrak m},
\underline{\mathfrak M}
\right):m_i=0,1,...,M_i-1,\:\mbox{for}\:\:i=1,...,d  \right \}.
\end{eqnarray}
Using these, we form the sets
\begin{eqnarray}
&&{\mathbb T^d}/{\mathcal K_1^{\underline{\mathfrak
N}}}=\left[0,\frac{1}{N_1}\right]
\times ...\times \left[0,\frac{1}{N_d}\right]\equiv
\Delta_1,\nonumber\\
&&{\mathbb T^d}/{\mathcal K_1^{\underline{\mathfrak
M}}}=\left[0,\frac{1}{M_1}\right]
\times ...\times \left[0,\frac{1}{M_d}\right]\equiv
\Delta_2\nonumber\\
&&\text{\rm and} \nonumber\\
&& Ann(\mathcal K_2^{\underline{\mathfrak M}})=
\left\{\gamma_{\underline{k}}(\underline{x})=e^{2\pi i\sum_{j=1}^d
M_jk_jx_j};
\quad \underline{k}=\left(k_1,...,k_d  \right)\in \mathbb Z^d \right
\}\; .\nonumber
\end{eqnarray}
Note that $\left\{\left(\prod_{j=1}^d M_i\right)^{\frac{1}{2}}
\gamma\right\}_{\gamma\in Ann(\mathcal K_2^{\underline{\mathfrak
M}})}$
is an  orthonormal  basis of $L^2(\Delta_2),$ and we have:

\begin{corollary}\label{cro4}
Let $g\in L^2(\mathbb T^d),$ and $N_i,M_i,\; i=1,2, \ldots, d$,  be
$2d$ positive integers such that:
\begin{eqnarray}
 B:&=&\frac{1}{\prod _{j+1}^d M_j} \;\sup_{\underline{x}\in \Delta_1 }
 \sum_{\left(\underline{\mathfrak m},\underline{\mathfrak M}\right)\in
 \mathcal K_2^{\underline{\mathfrak M}}}\mid
 \sum_{\left(\underline{\mathfrak n},\underline{\mathfrak N}\right)
 \in \mathcal K_1^{\underline{\mathfrak M}}}g
 \left(\left[\underline{x}-\left(\underline{\mathfrak n},
 \underline{\mathfrak N}\right)\right]\right)\nonumber\\
&\times&\overline{g\left(\left[\underline{x}-\left(\underline{\mathfrak
n},
\underline{\mathfrak N}\right)-\left(\underline{\mathfrak m},
\underline{\mathfrak M}\right)\right]\right)}\mid < \infty\; .
\end{eqnarray}
Then $\left\{ \gamma_{\underline k}T_{\left(\underline{\mathfrak n},
\underline{\mathfrak N}\right)}g
\right\}_{\left(\left(\underline{\mathfrak n},
\underline{\mathfrak N}\right),\underline{k}\right) \in
\mathcal K_1^{\underline{\mathfrak N}}\times \mathbb Z^d}$ is a Bessel
sequence for $L^2(\mathbb T^d)$ with upper bound $B$. If also
\begin{eqnarray}
&&A:=\frac{1}{\prod _{j+1}^d M_j} \inf_{\underline{x}\in \Delta_1
}\bigg[
\sum_{\left(\underline{\mathfrak n},\underline{\mathfrak N}\right)\in
\mathcal K_1^{\underline{\mathfrak M}}}\mid
g\left(\left[\underline{x}-
\left(\underline{\mathfrak n},\underline{\mathfrak
N}\right)\right]\right)
\mid^2\nonumber\\
&-&\sum_{\underline{\mathfrak o}\neq\left(\underline{\mathfrak m},
\underline{\mathfrak M}\right)\in \mathcal K_2^{\underline{\mathfrak
M}}}
\mid \sum_{\left(\underline{\mathfrak n},\underline{\mathfrak
N}\right)
\in \mathcal K_1^{\underline{\mathfrak M}}}g\left(\left[\underline{x}-
\left(\underline{\mathfrak n},\underline{\mathfrak
N}\right)\right]\right)
\overline{g\left(\left[\underline{x}-\left(\underline{\mathfrak n},
\underline{\mathfrak N}\right)-\left(\underline{\mathfrak m},
\underline{\mathfrak M}\right)\right]\right)}\mid \bigg]>0\;
,\nonumber\\
\end{eqnarray}
then $\left\{ \gamma_{\underline k}T_{\left(\underline{\mathfrak n},
\underline{\mathfrak N}\right)}g
\right\}_{\left(\left(\underline{\mathfrak n},
\underline{\mathfrak N}\right),\underline{k}\right) \in
\mathcal K_1^{\underline{\mathfrak N}}\times \mathbb Z^d}$  is a frame
for
$L^2(\mathbb T^d)$ with bounds $A,B.$
\end{corollary}

We observe immediately that the canonical commutation relations,
\begin{equation}
T_{\left(\underline{\mathfrak n},\underline{\mathfrak N}\right)}
\gamma_{\underline k}=e^{2\pi i\sum_{j=1}^d
M_j\left(\underline{\mathfrak n},
\underline{\mathfrak N}\right)_jk_j}\gamma_{\underline
k}T_{\left(\underline{\mathfrak n},
\underline{\mathfrak N}\right)}\;, \nonumber
\end{equation}
hold in this case. Indeed, for $g\in L^2(\mathbb T^d)$ we see that,
\begin{eqnarray}
\left(T_{\left(\underline{\mathfrak n},\underline{\mathfrak N}\right)}
\gamma_{\underline k}g\right)\left(\underline{x}\right)&=&
T_{\left(\underline{\mathfrak n},\underline{\mathfrak N}\right)}
\left(e^{2\pi i\sum_{j=1}^dM_jx_jk_j}g\left(\underline x
\right)\right)
=e^{2\pi i\sum_{j=1}^dM_j\left( x_j-\frac{n_j}{N_j} \right)k_j}
g\left(\left[\underline x -\left(\underline{\mathfrak n},
\underline{\mathfrak N}\right)\right]\right)\nonumber\\
&=&e^{-2\pi i\sum_{j=1}^d\frac{M_jn_jk_j}{N_j}}
\gamma_{\underline k}(\underline x)g\left(\left[\underline x -
\left(\underline{\mathfrak n},\underline{\mathfrak
N}\right)\right]\right)
\nonumber\\
&=&e^{-2\pi i\sum_{j=1}^dM_j\left(\underline{\mathfrak n},
\underline{\mathfrak N}\right)_jk_j}\gamma_{\underline k}(\underline
x)
T_{\left(\underline{\mathfrak n},\underline{\mathfrak N}\right)}g
\left(\underline x\right).\nonumber
\end{eqnarray}

It will be useful, for the purposes of the next section, to note that
the
frame operator for a frame
$\left\{ \gamma_{\underline k}T_{\left(\underline{\mathfrak n},
\underline{\mathfrak N}\right)}g
\right\}_{\left(\left(\underline{\mathfrak n},
\underline{\mathfrak N}\right),\underline{k}\right) \in
\mathcal K_1^{\underline{\mathfrak N}}\times \mathbb Z^d}$ commutes
with the
corresponding modulation and translation operators.

\begin{lemma}\label{Lem10}
Let $g\in L^2(\mathbb T^d)$  and let $N_i,M_i,\; i=1,2, \ldots, d$,  be
$2d$ positive
integers such that $\left\{ \gamma_{\underline
k}T_{\left(\underline{\mathfrak n},
\underline{\mathfrak N}\right)}g
\right\}_{\left(\left(\underline{\mathfrak n},
\underline{\mathfrak N}\right),\underline{k}\right) \in
\mathcal K_1^{\underline{\mathfrak N}}\times \mathbb Z^d}$ is a frame
for
$L^2(\mathbb T^d).$ If $S$ is the corresponding frame operator then,
$$ST_{\left(\underline{\mathfrak n_0},\underline{\mathfrak N}\right)}
\gamma_{\underline k_0}=T_{\left(\underline{\mathfrak n_0},
\underline{\mathfrak N}\right)}\gamma_{\underline k_0}S,\quad
\mbox{\rm for all}\quad \underline{k_0}\in \mathbb Z^d,\: \mbox{\rm and
all}\:
\left(\underline{\mathfrak n_0},\underline{\mathfrak N}\right)\in
\mathcal K_1^{\underline{\mathfrak N}}.$$
\end{lemma}

\begin{proof}
Let $f\in L^2(\mathbb T^d).$ We know that
\begin{eqnarray}
S(f)&=&\sum_{\underline {k} \in \mathbb
Z^d}\sum_{\left(\underline{\mathfrak n},
\underline{\mathfrak N}\right) \in \mathcal K_1^{\underline{\mathfrak
N}}}
\langle f\mid \gamma_{\underline k}T_{\left(\underline{\mathfrak n},
\underline{\mathfrak N}\right)}g\rangle \gamma_{\underline k}
T_{\left(\underline{\mathfrak n},\underline{\mathfrak N}\right)}g\;,
\end{eqnarray}
so that,
\begin{eqnarray}
\left(S\gamma_{\underline k_0}T_{\left(\underline{\mathfrak n_0},
\underline{\mathfrak N}\right)}\right)f &=&\sum_{\underline {k}
\in \mathbb Z^d}\sum_{\left(\underline{\mathfrak
n},\underline{\mathfrak N}\right)
\in \mathcal K_1^{\underline{\mathfrak N}}}\langle \gamma_{\underline
k_0}
T_{\left(\underline{\mathfrak n_0},\underline{\mathfrak
N}\right)}f\mid
\gamma_{\underline k}T_{\left(\underline{\mathfrak
n},\underline{\mathfrak N}
\right)}g\rangle \gamma_{\underline k}T_{\left(\underline{\mathfrak
n},
\underline{\mathfrak N}\right)}g\nonumber\\
&=&\sum_{\underline {k} \in \mathbb
Z^d}\sum_{\left(\underline{\mathfrak n},
\underline{\mathfrak N}\right) \in \mathcal K_1^{\underline{\mathfrak
N}}}
\langle f\mid T_{\left(-\underline{\mathfrak n_0},\underline{\mathfrak
N}\right)}
\gamma_{-\underline {k_0}}\gamma_{\underline
k}T_{\left(\underline{\mathfrak n},
\underline{\mathfrak N}\right)}g\rangle \gamma_{\underline k}
T_{\left(\underline{\mathfrak n},\underline{\mathfrak
N}\right)}g\nonumber\\
&=&\sum_{\underline {k} \in \mathbb
Z^d}\sum_{\left(\underline{\mathfrak n},
\underline{\mathfrak N}\right) \in \mathcal K_1^{\underline{\mathfrak
N}}}
\langle f\mid e^{2\pi i\sum_{j=1}^dM_j\left(\underline{\mathfrak n_0},
\underline{\mathfrak N}\right)_j(k_j-k_{0j})} \gamma_{\underline {k}-
\underline {k_0}}T_{\left(\underline{\mathfrak
{n-n_0}},\underline{\mathfrak N}\right)}
g\rangle \nonumber\\
&\times&\gamma_{\underline k}T_{\left(\underline{\mathfrak n},
\underline{\mathfrak N}\right)}g\nonumber\\
&=&\sum_{\underline {k} \in \mathbb
Z^d}\sum_{\left(\underline{\mathfrak n},
\underline{\mathfrak N}\right) \in \mathcal K_1^{\underline{\mathfrak
N}}}
\langle f\mid e^{2\pi i\sum_{j=1}^dM_j\left(\underline{\mathfrak n_0},
\underline{\mathfrak N}\right)_jk_j} \gamma_{\underline {k}}
T_{\left(\underline{\mathfrak {n}},\underline{\mathfrak N}\right)}g
\rangle \nonumber\\
&\times&\gamma_{\underline {k+k_0}}T_{\left(\underline{\mathfrak
n+n_0},
\underline{\mathfrak N}\right)}g\nonumber\\
&=&\sum_{\underline {k} \in \mathbb
Z^d}\sum_{\left(\underline{\mathfrak n},
\underline{\mathfrak N}\right) \in \mathcal K_1^{\underline{\mathfrak
N}}}
\langle f\mid e^{2\pi i\sum_{j=1}^dM_j\left(\underline{\mathfrak n_0},
\underline{\mathfrak N}\right)_jk_j} \gamma_{\underline {k}}
T_{\left(\underline{\mathfrak {n}},\underline{\mathfrak N}\right)}
g\rangle \nonumber\\
&\times&e^{2\pi i\sum_{j=1}^dM_j\left(\underline{\mathfrak n_0},
\underline{\mathfrak N}\right)_jk_j} \gamma_{\underline {k_0}}
T_{\left(\underline{\mathfrak n_0},\underline{\mathfrak N}\right)}
\gamma_{\underline {k}}T_{\left(\underline{\mathfrak n},
\underline{\mathfrak N}\right)}g\nonumber\\
&=&\sum_{\underline {k} \in \mathbb
Z^d}\sum_{\left(\underline{\mathfrak n},
\underline{\mathfrak N}\right) \in \mathcal K_1^{\underline{\mathfrak
N}}}
\langle f\mid  \gamma_{\underline {k}}T_{\left(\underline{\mathfrak
{n}},
\underline{\mathfrak N}\right)}g\rangle \gamma_{\underline {k_0}}
T_{\left(\underline{\mathfrak n_0},\underline{\mathfrak N}\right)}
\gamma_{\underline {k}}T_{\left(\underline{\mathfrak n},
\underline{\mathfrak N}\right)}g\nonumber\\
&=&\left(\gamma_{\underline {k_0}}T_{\left(\underline{\mathfrak n_0},
\underline{\mathfrak N}\right)}S\right)f.\nonumber
\end{eqnarray}\nopagebreak\end{proof}

\subsection{Necessary condition for having frames on the torus $\mathbb
T^d$}
We derive, in the next theorem, conditions for the existence of frames on $L^2(\mathbb
T^d)$,
which will be analogues of the conditions imposed by the product $ab$,
in Theorem \ref{Th0}, for
$\{E_{mb}T_{na}g  \}_{m,n\in \mathbb Z}$ to be a Gabor frame for $L^2(\mathbb
R)$.

We start by fixing a certain partition of  $\mathbb T^d.$
For $k=1,...,d$, let $i_k\in \{0,1,...,N_k-1  \}.$ For a $d-$tuplet
$(i_1,i_2,...,i_d)$,
of such $i_k$, let us define the subset $\Gamma _{(i_1,i_2,...,i_d)}$
by
\begin{eqnarray}
\Gamma
_{(i_1,i_2,...,i_d)}=\left[\frac{i_1}{N_1},\frac{i_1+1}{N_1}\right]
\times ....\times \left[\frac{i_d}{N_d},\frac{i_d+1}{N_d}\right]\; .
\end{eqnarray}
It is easy to see that these subsets have the following properties:
\begin{eqnarray}\label{c1}
\cup_{i_k=0,...,N_k-1; k=1,...,d}\Gamma _{(i_1,i_2,...,i_d)}=\mathbb
T^d\; ,
\end{eqnarray}

\begin{eqnarray}\label{c2}
&&\Gamma _{(i_1,i_2,...,i_d)}\cap \Gamma _{(i_1',i_2',...,i_d')}=
\emptyset\:  a.e.\: \mbox{if} \: (i_1,i_2,...,i_d)\neq
(i_1',i_2',...,i_d')\; ,
\end{eqnarray}

\begin{eqnarray}\label{c3}
&&T_{\left(\underline{\mathfrak n},\underline{\mathfrak N}\right)}
\left( \Gamma _{(i_1,i_2,...,i_d)} \right)\cap \Gamma
_{(i_1,i_2,...,i_d)}=
\emptyset, \; a.e.
\end{eqnarray}
for all $\left(\underline{\mathfrak o},\underline{\mathfrak N}\right)
\neq\left(\underline{\mathfrak n},\underline{\mathfrak N}\right)\in
\mathcal K_1^{\underline{\mathfrak N}}$ and  all
$(i_1,i_2,...,i_d)\;.$

  Let
  $$\mathcal F(\mathbb T^d)=\left\{f\in L^2(\mathbb T^d)\;:\;
\exists (i_1,i_2,...,i_d)\;\; \mbox{\rm and}\;\; supp(f)\subset
\Gamma _{(i_1,i_2,...,i_d)}\right\} $$
Then, by virtue of (\ref{c1}),
\begin{equation}
\mathcal F(\mathbb T^d)\quad \mbox{is dense in}\quad L^2(\mathbb T^d)\; ,
\end{equation}
and by (\ref{c3}),
\begin{equation}\label{IP}
\forall f\in \mathcal F(\mathbb T^d), f(\underline{x}).
\overline{T_{\left(\underline{\mathfrak n},\underline{\mathfrak
N}\right)}
f(\underline{x})}=0,
\end{equation}
for almost all $\underline{x}\in {\mathbb T^d}$ and all
$\left(\underline{\mathfrak n},\underline{\mathfrak N}
\right)\in \mathcal K_1^{\underline{\mathfrak N}}$ and
$\left(\underline{\mathfrak o},\underline{\mathfrak N}\right)\neq
\left(\underline{\mathfrak n},\underline{\mathfrak N}\right)$.

The following gives a necessary condition for having frame on
$L^2(\mathbb T^d)$
\begin{theorem}\label{Th5}
Let $g\in L^2(\mathbb T^d),$ and let $N_i,M_i,\; i=1,2, \ldots , d,$ be
$2d$ positive integers. Then the following hold:
\begin{enumerate}
\item[(i)]
$\mbox{If}\qquad {\left(\prod_{i=1}^d M_i\right)}>{\left(\prod_{j=1}^d
N_j\right)},
$\quad \mbox{then}

$\left\{ \gamma_{\underline k}T_{\left(\underline{\mathfrak
n},\underline{\mathfrak N}
\right)}g \right\}_{\left(\left(\underline{\mathfrak
n},\underline{\mathfrak N}\right),
\underline{k}\right) \in \mathcal K_1^{\underline{\mathfrak N}}\times
\mathbb Z^d}
\quad\mbox{is a not a frame for}\: L^2(\mathbb T^d).$
\item[(ii)]
$\mbox{If}\quad\left\{ \gamma_{\underline
k}T_{\left(\underline{\mathfrak n},
\underline{\mathfrak N}\right)}g
\right\}_{\left(\left(\underline{\mathfrak n},
\underline{\mathfrak N}\right),\underline{k}\right) \in
\mathcal K_1^{\underline{\mathfrak N}}\times \mathbb Z^d}\nonumber\:
\mbox{is a frame}\: for\: L^2(\mathbb T^d)$,\:then

$\prod_{i=1}^d M_i=\prod_{j=1}^d N_j\Leftrightarrow
\left\{\gamma_{\underline k}T_{\left(\underline{\mathfrak n},
\underline{\mathfrak N}\right)}g
\right\}_{\left(\left(\underline{\mathfrak n},
\underline{\mathfrak N}\right),\underline{k}\right) \in
\mathcal K_1^{\underline{\mathfrak N}}\times \mathbb Z^d}$ \mbox{is a
Riesz basis.}
\end{enumerate}
\end{theorem}

\begin{proof}
Let us assume that $\left\{\gamma_{\underline
k}T_{\left(\underline{\mathfrak n},
\underline{\mathfrak N}\right)}g
\right\}_{\left(\left(\underline{\mathfrak n},
\underline{\mathfrak N}\right),\underline{k}\right) \in
\mathcal K_1^{\underline{\mathfrak N}}\times \mathbb Z^d}$ is a frame
for
$L^2(\mathbb T^d)$ with frame operator $S$. Then
$\left\{S^{-\frac{1}{2}}
\left(\gamma_{\underline k}T_{\left(\underline{\mathfrak
n},\underline{\mathfrak N}
\right)}g \right)\right\}_{\left(\left(\underline{\mathfrak n},
\underline{\mathfrak N}\right),\underline{k}\right) \in
\mathcal K_1^{\underline{\mathfrak N}}\times \mathbb Z^d}$ is a tight
frame
for $L^2(\mathbb T^d)$  with frame bounds $1.$
Let $f\in\mathcal F(\mathbb T^d).$ Using $Lemma\:\ref{Lem3}$,
$Lemma\:\ref{Lem10},$
and $(\ref{IP}),$ we have:
\begin{eqnarray}
\int_{\mathbb T^d}\mid f(\underline{x})\mid^2 d\underline{x}&=&
\sum_{\underline {k} \in \mathbb Z^d}\sum_{\left(\underline{\mathfrak
n},
\underline{\mathfrak N}\right) \in \mathcal K_1^{\underline{\mathfrak
N}}}
\mid\langle f\mid S^{-\frac{1}{2}}\left( \gamma_{\underline k}
T_{\left(\underline{\mathfrak n},\underline{\mathfrak
N}\right)}g\right)
\rangle\mid^2 \nonumber\\
&=&\sum_{\underline {k} \in \mathbb
Z^d}\sum_{\left(\underline{\mathfrak n},
\underline{\mathfrak N}\right) \in \mathcal K_1^{\underline{\mathfrak
N}}}
\mid\langle f\mid \gamma_{\underline k}T_{\left(\underline{\mathfrak
n},
\underline{\mathfrak N}\right)}S^{-\frac{1}{2}}g\rangle\mid^2
\nonumber\\
&=&\frac{1}{\prod_{j=1}^d M_j}\int_{\mathbb T^d}\mid
f(\underline{x})\mid^2
\sum_{\left(\underline{\mathfrak n},\underline{\mathfrak N}\right)\in
\mathcal K_1^{\underline{\mathfrak M}}}\mid S^{-\frac{1}{2}}
g\left(\left[\underline{x}-\left(\underline{\mathfrak n},
\underline{\mathfrak N}\right)\right]\right)\mid^2dx,
\end{eqnarray}
which implies that
\begin{eqnarray}
\sum_{\left(\underline{\mathfrak n},\underline{\mathfrak N}\right)\in
\mathcal K_1^{\underline{\mathfrak M}}}\mid S^{-\frac{1}{2}}
g\left(\left[\underline{x}-\left(\underline{\mathfrak n},
\underline{\mathfrak N}\right)\right]\right)\mid^2=\prod_{j=1}^d M_j,
\quad a.e.,\: in\: \mathbb T^d.
\end{eqnarray}
Since
$$ \left\{S^{-\frac{1}{2}} \left(\gamma_{\underline k}
T_{\left(\underline{\mathfrak n},\underline{\mathfrak N}\right)}
g \right)\right\}_{\left(\left(\underline{\mathfrak n},
\underline{\mathfrak N}\right),\underline{k}\right) \in
\mathcal K_1^{\underline{\mathfrak N}}\times \mathbb Z^d}$$
is a tight frame, we have
\begin{eqnarray}
1&\geq& \parallel S^{-\frac{1}{2}} \left(\gamma_{\underline k}
T_{\left(\underline{\mathfrak n},\underline{\mathfrak N}\right)}
g \right)\parallel^2\nonumber\\
&=&\int_{\mathbb T^d}\mid S^{-\frac{1}{2}} g(\underline x)\mid^2
d{\underline x}\nonumber\\
&=&
\int_0^{\frac{1}{N_1}}\int_0^{\frac{1}{N_2}}...\int_0^{\frac{1}{N_d}}
\sum_{\left(\underline{\mathfrak n},\underline{\mathfrak N}\right)\in
\mathcal K_1^{\underline{\mathfrak M}}}\mid S^{-\frac{1}{2}}
g\left(\left[\underline{x}-\left(\underline{\mathfrak n},
\underline{\mathfrak N}\right)\right]\right)\mid^2 d{\underline
x}\nonumber\\
&=&
\int_0^{\frac{1}{N_1}}\int_0^{\frac{1}{N_2}}...\int_0^{\frac{1}{N_d}}
\prod_{j=1}^d M_jd{\underline x}\nonumber\\
&=& \left(\prod_{j=1}^d M_j\right)\left(\prod_{j=1}^d
N_j\right)^{-1},\nonumber
\end{eqnarray}
which proves (i).
In order to prove part (ii), let us assume that $\left\{
\gamma_{\underline k}
T_{\left(\underline{\mathfrak n},\underline{\mathfrak N}\right)}
g \right\}_{\left(\left(\underline{\mathfrak n},\underline{\mathfrak
N}\right),
\underline{k}\right) \in \mathcal K_1^{\underline{\mathfrak N}}\times
\mathbb Z^d}$
is  a Riesz Basis. Then $\left\{ \gamma_{\underline k}
T_{\left(\underline{\mathfrak n},\underline{\mathfrak N}\right)}
S^{-\frac{1}{2}}g \right\}_{\left(\left(\underline{\mathfrak n},
\underline{\mathfrak N}\right),\underline{k}\right) \in
\mathcal K_1^{\underline{\mathfrak N}}\times \mathbb Z^d}\nonumber
\mbox{is  a Riesz Basis}$ having bounds\\ $A=B=1,$ which implies that
$\parallel S^{-\frac{1}{2}}g\parallel^2=1.$ So we have
$\prod_{j=1}^dM_j=
\prod_{j=1}N_j.$ For the second implication, let us assume that
$\prod_{j=1}^dM_j=
\prod_{j=1}N_j.$ Then $\parallel \gamma_{\underline
k}T_{\left(\underline{\mathfrak n},
\underline{\mathfrak N}\right)}S^{-\frac{1}{2}}g  \parallel^2=1,$
\mbox{for all}
$\underline{k}\in Z^d$ \mbox{and all} $\left(\underline{\mathfrak n},
\underline{\mathfrak N}\right)\in \mathcal K_1^{\underline{\mathfrak
N}}\;.$
  Thus,
$\left\{ \gamma_{\underline k}T_{\left(\underline{\mathfrak n},
\underline{\mathfrak N}\right)}S^{-\frac{1}{2}}
g \right\}_{\left(\left(\underline{\mathfrak n},
\underline{\mathfrak N}\right),\underline{k}\right) \in
\mathcal K_1^{\underline{\mathfrak N}}\times \mathbb Z^d}$
is  a an orthonormal Basis for  $L^2(\mathbb T^d)$, and therefore,
\begin{equation}
\left\{ \gamma_{\underline k}T_{\left(\underline{\mathfrak n},
\underline{\mathfrak N}\right)}g
\right\}_{\left(\left(\underline{\mathfrak n},
\underline{\mathfrak N}\right),\underline{k}\right) \in
\mathcal K_1^{\underline{\mathfrak N}}\times \mathbb Z^d}=
\left\{S^{\frac{1}{2}} \gamma_{\underline
k}T_{\left(\underline{\mathfrak n},
\underline{\mathfrak N}\right)}S^{-\frac{1}{2}}g
\right\}_{\left(\left(\underline{\mathfrak n},\underline{\mathfrak N}
\right),\underline{k}\right) \in \mathcal K_1^{\underline{\mathfrak
N}}\times \mathbb Z^d}
\end{equation}
is a Riesz basis.
\end{proof}

\subsection{Generalization to arbitrary LCA group}
  We show next that the above  theorem can be extended to any LCA
group $G$, thus
constituting a generalization of Theorem $\ref{Th0}$ to any such group.
Before stating the
result, we note  two commutation relations.

For ${(k_1,\gamma_2)\in K_1\times Ann( K_2)},$ we have
\begin{eqnarray}
T_{k_1}\gamma_2=\gamma_2\left( k_1^{-1}\right)\gamma_2 T_{k_1}.
\end{eqnarray}
Also, for ${(k_1^0,\gamma_2^0)\in K_1\times Ann( K_2)},$  we have
\begin{equation}
S\gamma_2^0T_{k_1^0} = \gamma_2^0T_{k_1^0}S\; .
\end{equation}
Indeed, for $f,g\in L^2(G),$
\begin{eqnarray}
\left(S\gamma_2^0T_{k_1^0}\right)f&=&\sum_{k_1\in K_1}\sum_{\gamma_2\in
Ann(K_2)}
\langle
\gamma_2^0T_{k_1^0}f|\gamma_2T_{k_1}g\rangle\gamma_2T_{k_1}g\nonumber\\
&=&\sum_{k_1\in K_1}\sum_{\gamma_2\in Ann(K_2)}
\langle f|T_{{\left(k_1^0\right)}^{-1}}{\left(\gamma_2^0\right)}^{-1}
\gamma_2T_{k_1}g\rangle\gamma_2T_{k_1}g\nonumber\\
&=&\sum_{k_1\in K_1}\sum_{\gamma_2\in Ann(K_2)}\langle
f|\left({\left(\gamma_2^0\right)}^{-1}
\gamma_2\right)(k_1^0){\left(\gamma_2^0\right)}^{-1}
\gamma_2
T_{k_1{\left(k_1^0\right)}^{-1}}g\rangle\gamma_2T_{k_1}g\nonumber\\
&=&\sum_{\tilde k_1\in K_1}\sum_{\tilde\gamma_2\in Ann(K_2)}
\langle f|\tilde\gamma_2(k_1^0)\tilde\gamma_2T_{\tilde k_1}g
\rangle \tilde\gamma_2(k_1^0)\gamma_2^0T_{k_1^0}\tilde\gamma_2T_{\tilde
k_1}g\nonumber\; ,
\end{eqnarray}
whence the result.

\begin{theorem}\label{Th6}
Let $g\in L^2(G),$ and $K_1$ and $K_1$ be two uniform lattices in $G.$
Then,
the following hold:
\begin{enumerate}
\item[(i)]
$\mbox{If}\qquad \frac{\mid G/{K_1}\mid}{\mid G/{K_2}\mid} >1,$\quad
\mbox{then}

$\{\Theta^g_{k_1,\gamma_2 } \}_{(k_1,\gamma_2)\in K_1\times Ann( K_2)}
\quad\mbox{is a not a frame for}\: L^2(G).$
\item[(ii)]
$\mbox{If}\quad \{\Theta^g_{k_1,\gamma_2 } \}_{(k_1,\gamma_2)\in
K_1\times
Ann( K_2)}\nonumber\: \mbox{is a frame}\: for\: L^2(G)$,\:then

$\mid G/{K_1}\mid=\mid G/{K_2}\mid\Leftrightarrow
\{\Theta^g_{k_1,\gamma_2 } \}_{(k_1,\gamma_2)\in K_1\times Ann( K_2)}
$ \mbox{is a Riesz basis.}
\end{enumerate}
\end{theorem}

\begin{proof}
Let $\sigma:G/{\mathcal K_1}\longrightarrow G,$ be any section. For
any
$ k_1\in \ K_1,$ let $\sigma_{k_1}=
T_{k_1}\left( \sigma\left( G/{\mathcal K_1}  \right) \right).$ We
have:

 \begin{enumerate}
\item[(a)]
$\cup_{k_1\in \mathcal K_1}\sigma_{k_1}= G$
\item[(b)]
$\sigma_{k_1}\cap \sigma_{k_2}  =\emptyset\quad a.e.\quad in
\quad G,\quad \forall \:k_1\:\neq \:k_2.$
\end{enumerate}
Let
$$\mathcal F(G)=\left\{f\in L^2(G)\;:\;\exists k_1\in
\mathcal K_1\quad \mbox{\rm and}\quad supp(f)\subset \sigma_{k_1}
\right\}$$
and assume that $\{\Theta^g_{k_1,\gamma_2 } \}_{(k_1,\gamma_2)\in
\mathcal K_1\times Ann(\mathcal K_2)}$ is  a frame for
$L^2(G).$ Then, the set of vectors
$\{\Theta^{S^{-\frac{1}{2}}g}_{k_1,\gamma_2 } \}_{(k_1,\gamma_2)
\in\mathcal K_1\times Ann(\mathcal K_2)}$ is  a tight frame with bounds
1.
Let $f\in\mathcal F(G).$ Using Lemma\:\ref{Lem3}, and the statements
(a) and (b), we have:

\begin{eqnarray}
\int_{G}\mid f(\underline{x})\mid^2 d\underline{x}&=&\sum_{k_1
\in\mathcal K_1}\;\sum_{\gamma_2  Ann(\mathcal K_2)}\mid\langle
f\mid \gamma_2T_{k_1} S^{-\frac{1}{2}}g\rangle\mid^2 \nonumber\\
&=&\mid G/{K_2}\mid\int_{G}\mid f(\underline{x})\mid^2 \sum_{k_1
\in\mathcal K_1}\mid S^{-\frac{1}{2}}g\left(xk_1^{-1}\right)\mid^2dx,
\end{eqnarray}
which implies that
\begin{eqnarray}
\sum_{k_1 \in\mathcal K_1}\mid S^{-\frac{1}{2}}g\left(xk_1^{-1}\right)
\mid^2={\mid G/{K_2}\mid}^{-1},\quad \mbox{\rm a.e.\; in} \quad G.
\end{eqnarray}
Since $\left\{S^{-\frac{1}{2}} \left(\gamma_2T_{k_1}g \right)
\right\}_{(k_1,\gamma_2) \in\mathcal  K_1\times Ann(\mathcal K_2)}$
is a tight frame, we have
\begin{eqnarray}
1&\geq& \parallel S^{-\frac{1}{2}} \left(\gamma_2T_{k_1}g
\right)\parallel^2\nonumber\\
&=&\int_{G}\mid S^{-\frac{1}{2}} g(\underline x)\mid^2 d{\underline
x}\nonumber\\
&=& \int_{G/{\mathcal K_1}}\sum_{k_1 \in\mathcal K_1}
\mid S^{-\frac{1}{2}}g\left(xk_1^{-1}\right)\mid^2 d{\underline
x}\nonumber\\
&=& \int_{G/{\mathcal K_1}}{\mid G/{K_2}\mid}^{-1} d x\nonumber\\
&=& \mid G/{K_1}\mid.{\mid G/{K_2}\mid}^{-1}=\frac{\mid
G/{K_1}\mid}{\mid G/{K_2}\mid},
\nonumber
\end{eqnarray}
which proves (i). For part (ii) let assume that
$\left\{ \gamma_2T_{k_1}g \right\}_{(k_1,\gamma_2) \in\mathcal
K_1\times Ann(K_2)}
\nonumber \mbox{is  a Riesz basis.}$ Then the set of vectors
$\left\{S^{-\frac{1}{2}}
\left(\gamma_2T_{k_1}g \right)\right\}_{(k_1,\gamma_2) \in\mathcal
K_1\times Ann(\mathcal K_2)}$ is  a Riesz Basis having
bounds $A=B=1,$ which implies that $\parallel
S^{-\frac{1}{2}}g\parallel^2=1.$
So, we have $\mid G/{K_1}\mid=\mid G/{K_2}\mid.$ For the second
implication,
let assume that $\mid G/{K_1}\mid=\mid G/{K_2}\mid,$ then $\parallel
\gamma_2T_{k_1}S^{-\frac{1}{2}}g  \parallel^2=1,$ \mbox{for all}
${(k_1,\gamma_2) \in\mathcal  K_1\times Ann(\mathcal K_2)}.$  Then
$\left\{ \gamma_2T_{k_1}S^{-\frac{1}{2}}g \right\}_{(k_1,\gamma_2)
\in\mathcal  K_1\times Ann(\mathcal K_2)}\nonumber$ is  a an
orthonormal Basis for  $L^2(G),$ and therefore,
\begin{equation}
\left\{ \gamma_2T_{k_1}g \right\}_{(k_1,\gamma_2) \in
\mathcal K_1\times Ann(\mathcal K_2)}= \left\{S^{\frac{1}{2}}
\gamma_2T_{k_1}S^{-\frac{1}{2}}g \right\}_{(k_1,\gamma_2) \in
\mathcal K_1\times Ann(\mathcal K_2)}
\end{equation}
is a Riesz basis.
\end{proof}

\section{General shift-invariant systems}
In this section, using an obvious generalization of the notion of a
shift invariant system, defined in Section \ref{sec:prelim},
we present a complete characterization of a
generalized
Gabor frame on $L^2(G),$ where $G$ is any locally compact Abelian
group. The result
will be an extension of the result of Ron and Shen \cite{RS} on
$L^2(\mathbb R).$
Recall that for an LCA group the Fourier transform is a map
$\mathcal F$ defined from $L^1(G)\longrightarrow C(\hat G)$ by

\begin{equation}
(\mathcal F f)(\xi):= \hat f(\xi)= \int_G \overline{\langle
x,\xi\rangle}f(x)dx.
\end{equation}
This can be extended to a map
$L^2(G)\longrightarrow L^2(\hat G)$, satisfying the well-known
Plancherel identity.
The following properties of the Fourier transform will be required in
the sequel:
\begin{eqnarray}\label{f1}
\left(\widehat{T_yf}\right)(\xi)&=&\int_G \overline{\langle
x,\xi\rangle}
f(y^{-1}x)dx=\int_G \overline{\langle yx,\xi\rangle}f(x)dx\nonumber\\
&=& \overline{\langle y,\xi\rangle}\hat f(\xi),
\end{eqnarray}

\begin{eqnarray}\label{f2}
\left(\widehat{\eta f}\right) (\xi)&=&\int_G \overline{\langle
x,\xi\rangle}
\langle x,\eta\rangle f(x)dx= \hat f\left(\eta^{-1}\xi\right).
\end{eqnarray}

Let $\{g_m\}_{m\in \mathbb Z}$ be a collection of functions in $L^2(G)$
and
$K_1$ a uniform lattice in $G.$ For $m\in \mathbb Z$ and $k_1\in K_1,$
consider the function $g_{k_1,m}$ defined on $G$ by $g_{k_1,m}(x)=
g_m(xk_1^{-1}).$

\begin{lemma}\label{Lem11}
Let $\{  g_{k_1,m}\}_{m\in \mathbb Z; k_1\in K_1}$
and $\{  h_{k_1,m}\}_{m\in \mathbb Z; k_1\in K_1} $ be two shift
invariant systems and
assume that they are Bessel sequences. Then, for $e,\:f\in L^2(G)$, the
function,
\begin{eqnarray}
P\left(e,f \right): &G&\longrightarrow \mathbb C\nonumber\\
&x&\longmapsto \sum_{m\in \mathbb Z}\sum_{k_1\in K_1}\langle T_xe\mid
g_{k_1,m}
\rangle \langle h_{k_1,m}\mid T_xf\rangle
\end{eqnarray}
is continuous and well defined on $G/{K_1}.$ Its Fourier series in
$L^2\left(G/{K_1} \right)$ is
\begin{eqnarray}
P\left(e,f \right)(x)&=&\sum_{\gamma_1\in Ann(K_1)}c_{\gamma_1}
\gamma_1 (x)\; ,
\end{eqnarray}
where,
\begin{eqnarray*}
c_{\gamma_1}&=& {\mid G/{ K_1}\mid}^{-1}\int_{\hat G} \hat{e}(\xi)
\overline{\hat{f}(\xi\gamma_1)}\sum_{m\in \mathbb Z}
\overline{\hat{g_m}(\xi)}
\hat{h_i}(\xi\gamma_1)d\xi,\quad \gamma_1\in Ann(K_1).
\end{eqnarray*}
\end{lemma}

\begin{proof}
Using the Cauchy-Schwarz inequality and the fact that the sets  $\{
g_{k_1,m}\}_{m\in
\mathbb Z; k_1\in K_1}$ and $\{  h_{k_1,m}\}_{m\in \mathbb Z; k_1\in
K_1}$
are Bessel sequences, we conclude that the series defined by
$P\left(e,f \right)(x)$
converges absolutely. Also, for any $k_1\in K_1,$ we have $P\left(e,f
\right)(xk_1)=
P\left(e,f \right)(x)$, for almost all $x\:\in G$. Hence $P\left(e,f
\right)$ is well
defined as a function on $G/{K_1}.$ For the determination of the
Fourier coefficients,
let assume that $e,\:f$ are continuous and have compact supports. Then the
coefficients
$c_{\gamma_1}$, with respect to $\{\gamma_1(x) \}_{\gamma_1\in
Ann(K_1)}$, are given by
\begin{eqnarray}\label{coe}
c_{\gamma_1}&=&{\mid G/{
K_1}\mid}^{-1}\int_{G/{K_1}}\rho\left(e,f\right)(x)
\overline{\gamma_1(x)}dx\nonumber\\
&=&{\mid G/{ K_1}\mid}^{-1}\sum_{m\in \mathbb Z}\sum_{k_1\in
K_1}\int_{G/{K_1}}
\langle T_xe\mid g_m\left(.k_1^{-1}\right)\rangle\langle h_m\left(
.k_1^{-1} \right)
\mid T_xf\rangle \overline{\gamma_1(x)}dx\nonumber\\
&=&{\mid G/{ K_1}\mid}^{-1}\sum_{m\in \mathbb Z}\int_{G}\langle
T_xe\mid g_m\rangle
\langle h_m\mid T_xf\rangle \overline{\gamma_1(x)}dx\nonumber\\
&=&{\mid G/{ K_1}\mid}^{-1}\sum_{m\in \mathbb Z}\int_{G}\langle
T_xe\mid g_m\rangle
\overline{\langle T_xf\mid h_m\rangle \gamma_1(x)}dx\; .
\end{eqnarray}
For an arbitrary $\phi\in L^2(G),$
\begin{eqnarray}\label{i1}
&&\langle T_xe\mid \phi\rangle=\langle \mathcal FT_xe\mid \mathcal
F\phi\rangle=
\int_{\hat G}\overline{\xi(x)}\hat e(\xi)\overline{\hat \phi(\xi)}d\xi
=\mathcal F\left( \hat{e}.\overline{\hat{\phi}} \right)(x).
\end{eqnarray}
The last equality is  justified by identifying $G$ and $\hat{\hat G}$
and
adopting the natural dual pairing.

   Also, using (\ref{i1}) and (\ref{f1}), we have:
\begin{eqnarray}\label{i2}
\langle T_xf\mid h_m\rangle\gamma_1(x)&=&\gamma_1(x)\mathcal F\left(
\hat{f}.
\overline{\hat{h}_m} \right)(x)\nonumber\\
&=&\langle x,\gamma_1\rangle \mathcal F\left(
\hat{f}.\overline{\hat{h}_m}
\right)(x)=\overline{\langle x,\gamma_1^{-1}\rangle}\mathcal F\left(
\hat{f}.
\overline{\hat{h}_m} \right)(x)\nonumber\\
&=&\mathcal F\left(T_{\gamma_1^{-1}} \hat{f}.\overline{\hat{h}_m}
\right)(x).
\end{eqnarray}
Using (\ref{i1}) and (\ref{i2}) in (\ref{coe}), we have
\begin{eqnarray}
c_{\gamma_1}&=&{\mid G/{ K_1}\mid}^{-1}\sum_{m\in \mathbb Z}\int_{G}
\mathcal F\left( \hat{e}.\overline{\hat{g}_m} \right)(x)\mathcal F
\overline{\left(T_{\gamma_1^{-1}} \hat{f}.\overline{\hat{h}_m}
\right)}(x)dx\nonumber\\
&=&{\mid G/{ K_1}\mid}^{-1}\sum_{m\in \mathbb Z}\int_{\hat G}
\left(\hat{e}.\overline{\hat{g}_m}\right)(\xi)\left[\overline{T_{\gamma_1^{-1}}
\left(\hat{f}.\overline{\hat{h}_m}\right)}\right](\xi)d\xi\nonumber\\
&=&{\mid G/{ K_1}\mid}^{-1}\int_{\hat G}\hat{e}(\xi).
\overline{\hat{f}(\xi\gamma_1)} \sum_{m\in \mathbb Z}
\overline{\hat{g}_m}(\xi)}{\hat{h}_m(\xi\gamma_1)d\xi.
\end{eqnarray}
\end{proof}

  We proceed to characterize the frame properties of shift-invariant
systems for $L^2(G)$,  where $G$ is an arbitrary LCA group.
Let $\{g_m\}_{m\in \mathbb Z},$ be a collection of functions in
$L^2(G)$
and $K_1$ a uniform lattice in $G.$ For $\xi\in \hat G,$ consider the
matrix valued function $H(\xi)=\{ \hat g_{\gamma_1,m}(\xi)=
\hat g_m\left(\xi\gamma_1^{-1}  \right)  \}_{m\in \mathbb Z;
\gamma_1\in Ann(K_1)}.$

\begin{proposition}\label{Prop10}
Assume that the system $\{g_{\gamma_1,m}\}_{(\gamma_1,m)\in
Ann(K_1)\times \mathbb Z},$
has finite upper frame bound $B.$ Then, for almost all $\xi\in
\hat{G}$,
$H(\xi)$ defines
a bounded linear operator from $\ell^2(\mathbb Z)$ into
$\ell^2\left(Ann(K_1)\right)$
with operator norm $\leq \left(\mid G/{K_1}\mid B  \right)^{1/2}.$
Explicitly,
\begin{equation}\label{cond10}
 \sum_{\gamma_1\in Ann(K_1)}\mid\sum_{m\in \mathbb Z} \hat
{g}_m(\xi\gamma_1)
 \beta_m\mid^2\leq \mid G/{K_1}\mid B\parallel
\underline{\beta}\parallel^2,
\end{equation}
for almost all $\xi\in \hat{G}$
and  $\underline{\beta}\in \ell^2(\mathbb Z)$.
\end{proposition}

\begin{proof}
Since $\hat{G}/{Ann\left(K_1\right)}$ is compact, let $d\xi$ be the
Haar measure
on $\hat{G}/{Ann\left(K_1\right)}$  normalized so  that
$\mid \hat{G}/{Ann\left(K_1\right)}
\mid=\frac{1}{\mid G/{K_1}\mid}.$ Using the  fact that
$\widehat{\left(\frac{\hat{G}}{Ann\left(K_1\right)}\right)}=Ann\left(Ann(K_1)\right)=
K_1,$ we see that $K_1$ is an orthonormal basis of $L^2\left(
\hat{G}/{Ann\left(K_1\right)},
\mid G/{ K_1}\mid d\xi \right),$ where the action is defined in a
natural way by
$k_1(\xi):={\xi(k_1)}.$

Let $\alpha_{k_1,m}\neq 0$ for only finitely many
$\left(k_1,m\right)\in K_1\times
\mathbb Z,$ and let
\begin{eqnarray}
\alpha_m(\xi)&=& \sum_{k_1\in K_1}\alpha_{k_1,m}\overline{\xi(k_1)} =
\sum_{k_1\in K_1}\alpha_{k_1,m}\overline{k_1(\xi)}.
\end{eqnarray}
For any $\gamma_1\in Ann\left(K_1\right),$ we have
$\alpha_m(\xi\gamma_1)=
\alpha_m(\xi).$ Thus, $\alpha_m$ is well defined as a function on
$\hat{G}/{Ann\left(K_1\right)},$ and we have:
\begin{eqnarray}
&&\int_{\hat{G}/{Ann\left(K_1\right)}}\sum_{\gamma_1\in
Ann\left(K_1\right)}
\mid\sum_{m\in \mathbb
Z}\alpha_m(\xi)\hat{g}_m(\xi\gamma_1)\mid^2d\xi=\\
&&\int_{\hat{G}}\mid\sum_{m\in \mathbb
Z}\alpha_{m}(\xi)\hat{g}_m(\xi)\mid^2d\xi.
\nonumber
\end{eqnarray}
Using Parseval's theorem (or the fact that the Fourier transform is an
unitary
operator) and $(\ref{f1})$, we have:
\begin{eqnarray}\label{10a}
&&\int_{\hat{G}}\mid \sum_{k_1\in K_1;m\in \mathbb
Z}\alpha_{k_1,m}\xi(k_1)
\hat{g}_m(\xi)\mid^2d\xi=\parallel \sum_{k_1\in K_1;m\in \mathbb Z}
\alpha_{k_1,m}g_{k_1,m} \parallel^2.
\end{eqnarray}
Also, we have
\begin{eqnarray}\label{10b}
\parallel \sum_{k_1\in K_1;m\in \mathbb Z}\alpha_{k_1,m}g_{k_1,m}
\parallel^2&\leq& B \sum_{k_1\in K_1;m\in \mathbb
Z}\mid\alpha_{k_1,m}\mid^2,
\end{eqnarray}
and
\begin{eqnarray}\label{10c}
&&\parallel \underline{\alpha} \parallel^2= \sum_{k_1\in K_1;m\in
\mathbb Z}\mid\alpha_{k_1,m}\mid^2
=\mid G/{K_1}\mid\int_{\hat{G}/{Ann\left(K_1\right)}}\sum_{m\in
\mathbb Z}\mid \alpha_m(\xi) \mid^2d\xi.
\end{eqnarray}
Using $(\ref{10a})$, $(\ref{10b})$ and$(\ref{10c}),$ we obtain
\begin{eqnarray}
&&\int_{\hat{G}/{Ann\left(K_1\right)}}\sum_{\gamma_1\in
Ann\left(K_1\right)}
\mid\sum_{m\in \mathbb
Z}\alpha_m(\xi)\hat{g}_m(\xi\gamma_1^{-1})\mid^2d\xi\leq\\
&&\mid G/{K_1}\mid B \int_{\hat{G}/{Ann\left(K_1\right)}}\sum_{m\in
\mathbb Z}
\mid \alpha_m(\xi) \mid^2d\xi.\nonumber
\end{eqnarray}
For $\underline{\beta}\in \ell^2(\mathbb Z),$ with $\beta_m\neq 0$ for
only
finitely many $m\in \mathbb Z,$ choose $\alpha_m(\xi)=\beta_m\rho
(\xi)$,
where $\rho(\xi)= \sum_{k_1\in K_1}\rho_{k_1}k_1(\xi)$, with
$\rho_{k_1}\neq 0$
for only finitely many $k_1\in K_1$. Thus, we get
\begin{eqnarray}\label{10d}
&&\int_{\hat{G}/{Ann\left(K_1\right)}}\sum_{\gamma_1\in
Ann\left(K_1\right)}
\mid\rho(\xi)  \mid^2.\mid\sum_{m\in \mathbb
Z}\hat{g}_m(\xi\gamma_1^{-1})
\beta_m\mid^2d\xi\leq\\
&&\mid G/{K_1}\mid B \parallel \underline{\beta}
\parallel^2 \int_{\hat{G}/{Ann\left(K_1\right)}}\mid \rho(\xi)
\mid^2d\xi.\nonumber
\end{eqnarray}
Since the set of such $\rho$ is dense in $L^2\left(
\hat{G}/{Ann\left(K_1\right)}
\right)$ (because of the fact that $K_1$ is an orthonormal basis
$L^2\left( \hat{G}/{Ann\left(K_1\right)} \right)$), we have
\begin{eqnarray}\label{10e}
&&\sum_{\gamma_1\in Ann\left(K_1\right)}\mid\sum_{m\in
\mathbb Z}\hat{g}_m(\xi\gamma_1^{-1})\beta_m\mid^2 \leq\mid G/{K_1}
\mid B \parallel \underline{\beta} \parallel^2\; ,\nonumber
\end{eqnarray}
for almost all $\xi\in \hat{G}/{Ann\left(K_1\right)}$.
Let $V$ be a countable, dense subset of $\ell^2(\mathbb Z)$ of
$\underline{\beta}$'s with $\beta_m\neq 0$ for only finitely many
$m\in \mathbb Z,$ and let $N_1\subset \hat{G}/{Ann\left(K_1\right)}$
be the null set outside of which
\begin{eqnarray}\label{10f}
\sum_{\gamma_1\in Ann\left(K_1\right)}\mid\sum_{m\in \mathbb Z}
\hat{g}_m(\xi\gamma_1^{-1})\beta_m\mid^2 &\leq
&\mid G/{K_1}\mid B \parallel \underline{\beta}
\parallel^2,\:\forall\:
\underline{\beta}\in V.
\end{eqnarray}
Also, let $N_2\subset \hat{G}/{Ann\left(K_1\right)}$ be a null set
outside of which

\begin{eqnarray}\label{10g}
\sum_{m\in \mathbb Z}\mid\hat{g}_m(\xi\gamma_1^{-1})\mid^2 &\leq
&\mid G/{K_1}\mid .B,\:\forall\: \gamma_1\in Ann\left( K_1\right).
\end{eqnarray}
Letting $\underline{\beta}\in \ell^2(\mathbb Z)$ and
$\underline{\beta}^{(M)}\in V$ such that
$\underline{\beta}^{(M)}\longrightarrow
\underline{\beta}$, and applying Fatou's Lemma, we arrive at
\begin{eqnarray}\label{10h}
\sum_{\gamma_1\in Ann\left(K_1\right)}\mid\sum_{m\in \mathbb Z}
\hat{g}_m(\xi\gamma_1^{-1})\beta_m\mid^2 &\leq& \lim_{M\longrightarrow \infty} \inf \sum_{\gamma_1\in
Ann\left(K_1\right)}
\mid\sum_{m\in \mathbb
Z}\hat{g}_m(\xi\gamma_1^{-1})\beta_m^{(M)}\mid^2\nonumber\\
&\leq& \mid G/{K_1}\mid .B \parallel \underline{\beta} \parallel^2.
\end{eqnarray}
Finally, we have
\begin{eqnarray}\label{10i}
&&\sum_{\gamma_1\in Ann\left(K_1\right)}\mid\sum_{m\in
\mathbb Z}\hat{g}_m(\xi\gamma_1^{-1})\beta_m\mid^2 \leq
\mid G/{K_1}\mid .B \parallel \underline{\beta} \parallel^2\:
\forall\: \underline{\beta}\in \ell^2(\mathbb Z)\;, \nonumber
\end{eqnarray}
for almost all $\xi\in
\left\{\hat{G}/{Ann\left(K_1\right)}\right\}\setminus N$,
where $N=N_1\cup N_2$.
Since any element  $\nu\in\hat G$ can be written as $\nu=\xi\gamma_1,$
where
$\xi\in \hat{G}/{Ann\left(K_1\right)}$ and $\gamma_1\in
Ann\left(K_1\right)$
and since the first sum in $(\ref{10i})$ is taken over all elements of
$Ann\left(K_1\right),$ we have
\begin{eqnarray}\label{10j}
&&\sum_{\gamma_1\in Ann\left(K_1\right)}\mid\sum_{m\in
\mathbb Z}\hat{g}_m(\xi\gamma_1^{-1})\beta_m\mid^2 \leq
\mid G/{K_1}\mid. B \parallel \underline{\beta} \parallel^2;\:
\underline{\beta}\in \ell^2(\mathbb Z);\:\:a.e.\:\:\xi\in
\hat{G}.\nonumber
\end{eqnarray}
\end{proof}

    The following is a generalization of the Theorem \ref{SIS} to
$L^2(G),$
for any LCA group $G.$

\begin{theorem}\label{Th11} With the same setting as above, the
following hold:
\begin{enumerate}
\item[(i)]
$\{ g_{\gamma_1,m}\}$ is a Bessel sequence with upper bound $B$ if and
only if for almost all $\xi , \; H(\xi)$  defines a bounded operator
from
$\ell^2(\mathbb Z)$ into $\ell^2\left(Ann(K_1)\right)$ of norm
at most $\sqrt{\mid G/{K_1}\mid .B}.$
\item[(ii)]
$\{ g_{\gamma_1,m}\}$ is a frame for $L^2(G)$ with frame bounds $A,\:B$
if and only if
\begin{equation}\label{eq111}
\mid G/{K_1}\mid .A\mathbb I\leq H(\xi)H^*(\xi)\leq \mid G/{K_1}\mid
.B\mathbb I,
\end{equation}
for almost all $\xi$, where
$\mathbb I$ is identity operator on $\ell^2\left(Ann(K_1)\right).$
\item[(iii)]
$\{ g_{\gamma_1,m} \}$ is a tight frame for $L^2(G)$ if and only if
there is a
constant $c>0$ such that
\begin{equation}
\sum_{m\in \mathbb Z}\overline{\hat g_m(\xi)}\hat g_m(\xi\gamma_1)=
c\delta_{\gamma_1,1_G}\; , \quad \gamma_1\in Ann(K_1)\; ,
\end{equation}
for almost all $\xi$. In this case, the frame bound is $A=
\displaystyle\frac{c}{\mid G/{K_1}\mid}.$
\item[(iv)]
Two shift-invariant systems $\{ g_{\gamma_1,m} \}$ and $\{
h_{\gamma_1,m} \}$,
which form Bessel sequences, are dual frames if and only if
\begin{equation}\label{f6}
\sum_{m\in \mathbb Z}\overline{\hat g_m(\xi)}\hat h_m(\xi\gamma_1)=\mid
G/{K_1}
\mid\delta_{\gamma_1,1_{\hat G}}\; ,\quad \gamma_1\in Ann(K_1)\; .
\end{equation}
for almost all $\xi$.
\end{enumerate}
\end{theorem}

\begin{proof}
For part (iv), it's known that $\{ g_{\gamma_1,m} \}$ and $\{
h_{\gamma_1,m} \}$
are dual frames if and only if
\begin{equation}\label{CRB}
\langle e\mid f\rangle = \sum_{m\in \mathbb Z}\sum_{k_1\in
Ann(K_1)}\langle e\mid
g_{\gamma_1,m}\rangle\langle  h_{\gamma_1,m}\mid f\rangle,\quad \forall
e,\:f
\in \:L^2(G),
\end{equation}
and we have $\rho\left(e, f \right)(x)=\langle T_xe\mid T_xf\rangle=
\langle e\mid f\rangle,\quad \forall x\in G/{K_1}.$ Hence the functions
$\rho\left(e,
f \right)(x)$ and $\langle e\mid f\rangle$ have the same Fourier
coefficients in
$L^2\left(G/{K_1} \right)$, whence
\begin{eqnarray}\label{eq1}
c_{\gamma_1}&=&\frac{1}{\mid G/{K_1}\mid}\int_{\hat G}\hat{e}(\xi).
\overline{\hat{f}(\xi\gamma_1)} \sum_{m\in \mathbb Z}
\overline{\hat{g}_m}(\xi)}{\hat{h}_m(\xi\gamma_1)d\xi\nonumber\\
&=&\langle e\mid f\rangle\delta_{\gamma_1,1_{\hat G}}=
\delta_{\gamma_1,1_{\hat G}}\int_{\hat
G}\hat{e}(\xi)\overline{\hat{f}(\xi)}d\xi.
\end{eqnarray}
Since $(\ref{eq1})$ holds for all $e\in L^2\left(G\right),$  we have
\begin{eqnarray}\label{eq2}
\overline{\hat{f}(\xi\gamma_1)} \sum_{m\in \mathbb Z}
\overline{\hat{g}_m}(\xi)}{\hat{h}_m(\xi\gamma_1)&=&\mid G/{K_1}
\mid\delta_{\gamma_1,1_{\hat G}}\overline{\hat{f}(\xi)},\:
a.e.\:\xi\in\hat G.
\end{eqnarray}
If $\gamma_1=1_{\hat G},$ we get
\begin{eqnarray}
 \sum_{m\in \mathbb Z}\overline{\hat{g}_m}(\xi)}{\hat{h}_m(\xi)&=&\mid
G/{K_1}
 \mid,\quad a.e.\:\xi\in\hat G.
\end{eqnarray}
\mbox{If} $\gamma_1\neq 1_{\hat G},$ then $\gamma_1\xi\neq \xi,\quad
\forall
\xi\in \hat G.$ Since $\hat G$ is Hausdorff, there exists an open
neighbourhood,
$\mathcal O_{\gamma_1\xi}$ of $\gamma_1\xi$, such that
$\xi\notin \mathcal O_{\gamma_1\xi}.$ By taking $\hat f(s)=
\chi_{\mathcal O_{\gamma_1\xi}}(s),$ the charactiristic function of
$\mathcal O_{\gamma_1\xi},$
and using this function in (\ref{eq2}), we obtain
\begin{eqnarray}
 \sum_{m\in \mathbb
Z}\overline{\hat{g}_m}(\xi)}{\hat{h}_m(\xi\gamma_1)&=
 &0,
\end{eqnarray}
for almost all $\xi\in\hat G$ and for all $\gamma_1\neq 1_{\hat G}$.
Thus,
\begin{equation}
\sum_{m\in \mathbb Z}\overline{\hat{g}_m}(\xi)\hat{h}_m(\xi\gamma_1)=
\mid G/{K_1}\mid\delta_{\gamma_1,1_{\hat G}},\; \quad
\gamma_1\in Ann(K_1),
\end{equation}
for almost all $\xi\in \hat G$.
For the opposite implication let us assume that (\ref{f5}) holds. It
follows
that the function $\rho\left(e,f\right)$ and $\langle e\mid f\rangle$
have
the same Fourier coefficients and then $\rho\left(e,f\right)(x)=\langle
e\mid
f\rangle,\:\forall x\in G.$ By taking $x=1_G,$ we obtain
$(\ref{CRB}).$

 For the first part of (iii), we use (iv) and Lemma \ref{RB}. For the
 second part, we use (ii) and the fact that the $\gamma_1$-st row and
 $\gamma_2$-nd column of $H(\xi)H^*(\xi)$ are $\sum_{m\in \mathbb Z}
 \hat{g}_m (\xi\gamma_1^{-1})\overline{\hat{g}_m(\xi\gamma_2^{-1})}=
 c\delta_{\gamma_1,\gamma_2}$, to get $A= \frac{c}{\mid G/{K_1}
\mid}$.

For  part (ii), if $\sum_{\gamma_1\in Ann(K_1),m\in \mathbb Z}
\mid\langle f,g_{\gamma_1,m}\rangle\mid^2 \leq B\parallel
f\parallel^2,$
then  by virtue of Proposition \ref{Prop10}, $H(\xi)H^*(\xi)\leq \mid
G/{K_1}\mid B\mathbb I$. Let $f$ be an element in $L^2(G)$ with $\hat
f$
compactly supported and let $m\in \mathbb Z.$ The series
\begin{equation}\label{eq95}
\sum_{\gamma_1\in Ann(K_1)}\hat g_m(\xi \gamma_1)\hat f^*(\xi\gamma_1)
\end{equation}
defines a function in $L^2\left({\hat G}/{Ann(K_1)}  \right).$
Since $K_1$ is an orthogonal basis of $L^2\left({\hat G}/{Ann(K_1)}
\right)$,
it follows that
\begin{equation}\label{eq96}
\sum_{\gamma_1\in Ann(K_1)}\hat g_m(\xi \gamma_1)\hat
f^*(\xi\gamma_1)=
\sum_{k_1\in K_1}c_{k_1,m}\xi(k_1),
\end{equation}
where
\begin{eqnarray}\label{eq97}
c_{k_1,m}&=& \mid G/{K_1}\mid\int_{{\hat G}/{Ann(K_1)}}\sum_{\gamma_1
\in Ann(K_1)}\overline{\xi(k_1)}\hat g_m(\xi \gamma_1)
\hat f^*(\xi\gamma_1)d\xi\nonumber\\
&=&\mid G/{K_1}\mid\int_{{\hat G}}\overline{\xi(k_1)}\hat g_m(\xi)
\hat f^*(\xi)d\xi\nonumber\\
&=&\mid G/{K_1}\mid\int_{{\hat G}}\mathcal{F}\left(T_{k_1}g_m
\right)(\xi)\mathcal{F}\left( f^*\right)(\xi)d\xi\nonumber\\
&=&\mid G/{K_1}\mid\int_{{\hat G}}g_m(\xi {k_1}^{-1})
f^*(\xi)d\xi\nonumber\\
&=&\mid G/{K_1}\mid\langle g_{k_1 ,m},f  \rangle\; .
\end{eqnarray}
Thus,
\begin{equation}\label{eq98}
\int_{{\hat G}/{Ann(K_1)}}\mid \sum_{\gamma_1\in Ann(K_1)}\hat g_m(\xi
\gamma_1)\hat f^*(\xi\gamma_1)\mid ^2d\xi=\mid G/{K_1}\mid \sum_{k_1\in
K_1}
\mid \langle g_{k_1 ,m},f  \rangle  \mid^2.
\end{equation}
Let $\xi\in \hat G$ and set $\hat{\underline{f}}(\xi)=
\{\underline{f}(\xi\gamma_1^{-1})  \}_{\gamma_1\in K_1}.$
We then have
\begin{eqnarray}\label{eq99}
\int_{{\hat G}/{Ann(K_1)}}\parallel \underline{\hat{f}}(\xi)
\parallel^2 d\xi&=
&\int_{{\hat G}/{Ann(K_1)}}\sum_{\gamma_1\in Ann(K_1)}\mid
\hat f (\xi\gamma_1^{-1})\mid^2 d\xi\nonumber\\
&=&\int_{\hat G}\mid \hat f(\xi)\mid^2 d\xi\nonumber\\
&=&\parallel f\parallel^2\;,
\end{eqnarray}
while use of  $(\ref{eq98})$ yields
\begin{eqnarray}\label{eq100}
&&\int_{{\hat G}/{Ann(K_1)}}\parallel H^*(\xi)\underline{\hat{f}}(\xi)
\parallel^2d\xi\nonumber\\
&=&\sum_{m\in \mathbb Z} \int_{{\hat G}/{Ann(K_1)}}\mid \sum_{\gamma_1
\in Ann(K_1)}\hat g_m(\xi \gamma_1)\hat f^*(\xi\gamma_1)\mid
^2d\xi\nonumber\\
&=&{\mid G/{K_1}\mid}\sum_{m\in \mathbb Z}\sum_{k_1\in K_1}\mid
\langle g_{k_1 ,m},f  \rangle\mid^2.
\end{eqnarray}

Let $\hat\rho\in L^2 \left({\hat G}/{Ann(K_1)}  \right)$ and let
$\underline{\beta}\in \ell (\mathbb Z)$ with $\beta_k\neq 0$ for only
finitely many $k\in \mathbb Z.$
Take any section $\sigma:{\hat G}/{Ann(K_1)}\longrightarrow \hat G$
and let $\hat\rho_\sigma$ be the function defined on $\hat G$ with
support in
$V_\sigma=\sigma\left({\hat G}/{Ann(K_1)}\right)$ and such that
$\hat\rho_\sigma(\xi)=
\hat\rho(\pi(\xi))$ if $\xi\in V_\sigma$ and $\hat\rho_\sigma(\xi)=0$
otherwise.
It is then easy to see that
\begin{equation}
\int_{{\hat G}/{Ann(K_1)}}\mid \hat\rho(\xi)
\mid^2d\xi=\int_{V_\sigma}\mid
\hat\rho_\sigma(\xi) \mid^2d\xi.
\end{equation}
Assuming that $\mid Ann(K_1) \mid=\infty,$ let us write $Ann(K_1)$ in
the form
\begin{equation}
Ann(K_1)=\left\{ \gamma_{1,k}:k\in\mathbb Z  \right\}.
\end{equation}
Define a function $\hat f$ on $\hat G$ by $\hat f(\xi)=\beta_k
\hat\rho_\sigma(\xi\gamma_{1,k}),$ where $k$ is such that
$\xi\gamma_{1,k}\in V_\sigma$
and let $\underline {\hat f}(\xi)=\{\beta_k\hat
f(\xi\gamma^{-1}_{1,k})\}_{k\in
\mathbb Z}.$ It is then easy to see that
\begin{equation}
\underline {\hat f}(\xi)= \underline{\beta}\hat\rho_\sigma(\xi)\; .
\end{equation}
Furthermore, from $(\ref{eq99})$ and $(\ref{eq100})$ and the first
$(\ref{eq111})$, we have
\begin{eqnarray}\label{eq101}
&&\int_{{\hat G}/{Ann(K_1)}}\parallel H^*(\xi)\underline{\hat{f}}(\xi)
\parallel^2d\xi\\
&=&\int_{{\hat G}/{Ann(K_1)}}\mid \hat\rho\left(\xi\right)
\mid^2\parallel H^*(\xi)\underline{\beta} \parallel^2d\xi\nonumber\\
&=&{\mid G/{K_1}\mid}.\sum_{m\in \mathbb Z}\sum_{k_1\in K_1}
\mid\langle g_{k_1 ,m},f  \rangle\mid^2\nonumber\\
&\geq& \mid G/{K_1}\mid. A\parallel f\parallel^2\nonumber\\
&=& \mid G/{K_1}\mid. A\int_{{\hat G}/{Ann(K_1)}}\parallel
\underline{ f}(\xi)d\xi \parallel^2\nonumber\\
&=&\mid G/{K_1}\mid .A\int_{{\hat G}/{Ann(K_1)}}\sum_{k\in
\mathbb Z}\mid\beta_k.\hat\rho(\xi)\mid^2d\xi\nonumber\\
&=&\mid G/{K_1}\mid .A\parallel \underline{\beta}
\parallel^2\int_{{\hat G}/{Ann(K_1)}}\mid \hat\rho(\xi)
\mid^2d\xi.\nonumber
\end{eqnarray}
Letting $\hat\rho$ run over all of $L^2 \left({\hat
G}/{Ann(K_1)}\right)$,
we obtain
\begin{equation}\label{eq01}
\parallel H^*(\xi)\underline{\beta} \parallel^2\geq \mid G/{K_1}\mid.
A\parallel
\underline{\beta} \parallel^2\; ,
\end{equation}
for almost all $\xi\in {\hat G}/{Ann(K_1)}$,
where the null set involved in $(\ref{eq01})$ may depend on
$\underline\beta.$
Let $V$ be a countable dense set of $\underline\beta$'s in
$\ell^2(\mathbb Z)$
such that $\beta_k\neq 0$ for only finitely many $k\in \mathbb Z$
and
let $N_1\subset {\hat G}/{Ann(K_1)}$ be a null set such that
\begin{equation}\label{eq02}
\parallel H^*(\xi)\underline{\beta} \parallel^2\geq \mid G/{K_1}\mid
.A\parallel
\underline{\beta} \parallel^2;\:\beta\in V,  \:\xi\in {\hat
G}/{Ann(K_1)}/N_1.
\end{equation}
Also, let $N_2\subset {\hat G}/{Ann(K_1)}$ be a null set such that
\begin{equation}\label{eq03}
\parallel H^*(\xi)\underline{\beta} \parallel^2\leq \mid
G/{K_1}\mid.B\parallel
\underline{\beta} \parallel^2;\:\underline\beta\in \ell^2(\mathbb Z),
\:\xi\in {\hat G}/{Ann(K_1)}/N_2.
\end{equation}
Take $\xi\in \left({\hat G}/{Ann(K_1)}\right)-\left(N_2\cup
N_2\right)$,
$\underline\beta\in \ell^2(\mathbb Z)$ and $\underline\beta^{(M)}\in V$
such that
$\beta^{(M)}\longrightarrow \underline\beta.$ Then,
from $(\ref{eq02})$ and $(\ref{eq03})$, we conclude that
\begin{eqnarray}
\parallel H^*(\xi){\underline\beta} \parallel^2&=& \lim_{M\rightarrow
\infty}\parallel H^*(\xi){\underline\beta^{(M)}} \parallel^2\geq
\mid G/{K_1}\mid .A\lim_{M\rightarrow \infty}
\parallel{\underline\beta^{(M)}}
\parallel^2  \nonumber\\
&=&\mid G/{K_1}\mid .A\parallel{\underline\beta} \parallel^2.
\end{eqnarray}
This completes the proof of the implication ``$\Rightarrow$'' of part
(ii).
To prove the opposite implication, let $f\in L^2\left(G \right)$
such that $\hat f$ is compactly
supported in $\hat G.$ Then $(\ref{eq99})$ and $(\ref{eq100})$, imply
$$A\parallel f\parallel^2\leq\sum_{\gamma_1\in Ann(K_1),m\in \mathbb
Z}
\mid\langle f,g_{\gamma_1,m}\rangle\mid^2 \leq B\parallel
f\parallel^2.$$
\end{proof}

\section{Conclusion}We have concentrated in the present paper on
working  out
the essential mathematical results, generalizing some earlier
theoretical
results based on $G = \mathbb R^d$.  The cases $G = \mathbb R^d$ and
$G= \mathbb T^d$ are of
immediate physical interest. In a future publication, we intend to
dwell upon
some physical applications, which make use of coherent states
and frames built out of
the generalized Weyl-Heisenberg groups discussed here. In particular,
coherent
states and frames built on the cotangent bundle of the torus in
$d$-dimensions
could find useful applications in atomic physics.
While the  $p$-adic groups are also LCA groups,
we are unaware of any obvious physical applications of frames or
coherent
states built out of them. Also, since the Haar measure
plays a crucial role in the development of the results presented above,
we do not see
any immediate way  of extending these results to infinite dimensional
groups. In a
subsequent publication we also intend to present results generalizing
the Walnut representation \cite{CCJ} of the frame operator and its representation in frequency domain to the
present setting.

\section*{Acknowledgments}
The authors would like to thank H. Feichtinger, G. Kutinyok and Z. Shen
for useful
discussions. They also benefitted from grants from the  Natural
Sciences and Engineering
Research Council (NSERC) of Canada and the Fonds de recherche sur la
nature et les
technologies (FQRNT), Qu\'ebec.

\end{document}